\documentclass[%
 reprint,
superscriptaddress,
preprintnumbers,
 amsmath,amssymb,
 aps,
 prl,
]{revtex4-2}

\usepackage{graphicx}
\usepackage{dcolumn}
\usepackage{bm}
\usepackage{physics}
\usepackage{xcolor}
\usepackage{comment}
\usepackage{soul}

\newcommand{\mpl}{M_{\mathrm{pl}}}
\newcommand{\bp}{\mathbf{p}}

\newcommand{\bk}{\mathbf{k}}
\newcommand{\md}{\mathrm{d}}
\newcommand{\bz}{\bm{\zeta}}
\begin{document}

\preprint{RESCEU-20/22}

\title{Constraining Primordial Black Hole Formation from Single-Field Inflation}

\author{Jason Kristiano}
\email{jkristiano@resceu.s.u-tokyo.ac.jp}
\affiliation{Research Center for the Early Universe (RESCEU), Graduate School of Science, The University of Tokyo, Tokyo 113-0033, Japan}
\affiliation{Department of Physics, Graduate School of Science, The University of Tokyo, Tokyo 113-0033, Japan \looseness=-1}

\author{Jun'ichi Yokoyama}
\email{yokoyama@resceu.s.u-tokyo.ac.jp}
\affiliation{Kavli Institute for the Physics and Mathematics of the Universe (Kavli IPMU), WPI, UTIAS, The University of Tokyo, Kashiwa, Chiba 277-8568, Japan}
\affiliation{Research Center for the Early Universe (RESCEU), Graduate School of Science, The University of Tokyo, Tokyo 113-0033, Japan}
\affiliation{Department of Physics, Graduate School of Science, The University of Tokyo, Tokyo 113-0033, Japan \looseness=-1}
\affiliation{Trans-Scale Quantum Science Institute, The University of Tokyo, Tokyo 113-0033, Japan}


\date{\today}

\begin{abstract}
The most widely studied formation mechanism of a primordial black hole is collapse of large-amplitude  perturbation on small scales generated in single-field inflation. In this Letter, we calculate one-loop correction to the large-scale power spectrum in a model with sharp transition of the second slow-roll parameter. We find that models producing an appreciable amount of primordial black holes induce nonperturbative coupling on a large scale probed by cosmic microwave background radiation. Our result implies that a small-scale power spectrum can be constrained by large-scale cosmological observations.
\end{abstract}

\maketitle

Primordial black holes (PBHs) have been a research interest for more than 50 years \cite{Zel:1967, Hawking:1971ei, Carr:1974nx}, although there has been no observational evidence for them. They could be light enough for Hawking radiation to be important \cite{Hawking:1974rv}, they are a potential dark matter candidate \cite{Chapline:1975ojl, Garcia-Bellido:1996mdl, Ivanov:1994pa, Yokoyama:1995ex, Afshordi:2003zb, Frampton:2010sw, Belotsky:2014kca, Carr:2016drx, Inomata:2017okj, Espinosa:2017sgp} (reviewed in \cite{Green:2020jor, Carr:2020xqk}), and they can explain the origin of gravitational wave events \cite{Sasaki:2016jop, Raidal:2017mfl, Ali-Haimoud:2017rtz, Raidal:2018bbj, Vaskonen:2019jpv, Hall:2020daa}.

A number of formation mechanisms of PBHs in the early Universe have been proposed. The most popular one makes use of quantum fluctuations \cite{ Mukhanov:1981xt,Hawking:1982cz,Guth:1982ec, Starobinsky:1982ee} generated in cosmic inflation \cite{Starobinsky:1980te, Sato:1980yn, Guth:1980zm}. Observations of cosmic microwave background (CMB) anisotropy \cite{Planck:2018nkj, Planck:2018jri, Planck:2019kim} tightly constrain these fluctuations on large scales.  On smaller scales that cannot be probed by CMB observations, observational constraints are loose enough \cite{Nakama:2014vla, Jeong:2014gna, Inomata:2016uip, Nakama:2017qac, Kawasaki:2021yek, Kimura:2021sqz, Wang:2022nml}.  Therefore, it is possible to have a theory which produce large fluctuations with the amplitude of power spectrum $\mathcal{O}(0.01)$ that satisfy observational constraints
and many models have been proposed to realize such a feature 
\cite{Ivanov:1994pa, Carr:1993aq, Yokoyama:1995ex, Yokoyama:1998pt, Kawasaki:1997ju, Kawasaki:1998vx, Kawasaki:2016pql, Gao:2018pvq, Nanopoulos:2020nnh, Wu:2021zta, Stamou:2021qdk, Cheng:2018yyr, Ballesteros:2019hus, Pi:2017gih, Dalianis:2019asr, Dalianis:2018frf, Mahbub:2019uhl, Cicoli:2018asa, Ozsoy:2018flq, Cicoli:2022sih, Ezquiaga:2017fvi, Ballesteros:2017fsr, Drees:2019xpp, Cheong:2019vzl, Rasanen:2018fom, Garcia-Bellido:2017mdw, Ragavendra:2020sop, Mishra:2019pzq, Atal:2019cdz, Zheng:2021vda, Wang:2021kbh, Rezazadeh:2021clf, Iacconi:2021ltm, Cai:2021zsp, Kefala:2020xsx, Inomata:2021uqj, Ng:2021hll, Motohashi:2019rhu, Inomata:2021tpx, Hertzberg:2017dkh, Ballesteros:2020qam, Kannike:2017bxn, Di:2017ndc, Saito:2008em, Bugaev:2008bi, Solbi:2021wbo, Solbi:2021rse, Teimoori:2021pte, Ballesteros:2018wlw, Ballesteros:2021fsp, Ashoorioon:2019xqc, Frolovsky:2022ewg, Fu:2019ttf, Heydari:2021gea, Kawai:2021edk, Ozsoy:2020kat}.
Peaks of such fluctuations may collapse into PBHs with appreciable abundance \cite{Carr:2009jm,Carr:2020gox} after entering the horizon, which also produce large stochastic gravitational wave background \cite{Assadullahi:2009jc, Baumann:2007zm, Saito:2008jc, Saito:2009jt, Cai:2018dig} that can be probed by future gravitational wave observations as well as pulsar timing array experiments \cite{Yokoyama:2021hsa}.

The simplest inflation model that is consistent with current observational data \cite{Planck:2018jri, Planck:2019kim} is canonical slow-roll (SR) inflation (reviewed in \cite{Sato:2015dga}). It is described by a scalar field $\phi$, called inflaton, with a canonical kinetic term and potential $V(\phi)$ in quasi-de Sitter space. The standard SR inflation generates nearly scale-invariant adiabatic curvature perturbation that behaves classically as the decaying mode decreases exponentially during inflation, so that the perturbation variable and its conjugate momentum practically commute with each other \cite{Polarski:1995jg}. In order to be consistent with CMB observations \cite{Planck:2018jri, Planck:2019kim}, the shape of the potential is tightly constrained for a finite range of $\phi$.

If the inflaton passes through  an extremely flat region of the potential with $\md V/\md\phi \approx 0$ after the comoving scales probed by CMB have left the horizon, it may produce large-amplitude  fluctuations on small scales. In this region, SR condition fails, and the inflation goes into a temporary ultraslow-roll (USR) period \cite{Kinney:1997ne, Inoue:2001zt, Kinney:2005vj, Martin:2012pe, Motohashi:2017kbs}. During this regime the nonconstant mode of fluctuations, which would decay exponentially in SR inflation, actually grows, as observed in other models \cite{Yokoyama:1998rw,Saito:2008em}, resulting in enhanced power spectrum on specific scales. This may also imply the importance of quantum effects as we will see below. 
 
Many inflation models with a flat region or inflection point of the potential have been proposed inspired by high energy theories such as supergravity \cite{Kawasaki:1997ju, Kawasaki:1998vx, Kawasaki:2016pql, Gao:2018pvq, Nanopoulos:2020nnh, Wu:2021zta, Stamou:2021qdk}, axion monodromy \cite{Cheng:2018yyr, Ballesteros:2019hus}, scalaron in $R^2$-gravity \cite{Pi:2017gih}, $\alpha$-attractor \cite{Dalianis:2019asr, Dalianis:2018frf, Mahbub:2019uhl}, and string theory \cite{Cicoli:2018asa, Ozsoy:2018flq, Cicoli:2022sih}, as well as in Higgs inflation which does not require theories beyond the standard model \cite{Ezquiaga:2017fvi, Ballesteros:2017fsr, Drees:2019xpp, Cheong:2019vzl, Rasanen:2018fom}. As an extension of USR period, 
constant-roll inflation can also produce large amplitudes \cite{Motohashi:2014ppa, Motohashi:2019rhu}.

Theoretically, the power spectrum is described by the vacuum expectation value (VEV) of the fluctuation two-point functions in quantum field theory, to which only wavevectors with equal magnitude and opposite direction contribute.
As we expand the theory to higher order in fluctuations, we will get higher-order interaction terms, which generate primordial non-Gaussianity \cite{Maldacena:2002vr, Chen:2006nt, Seery:2005wm}. They  also generate back reaction to the two-point function which is called loop correction \cite{Weinberg:2005vy, Sloth:2006az, Seery:2007wf, Bartolo:2010bu, Senatore:2009cf, delRio:2018vrj, Melville:2021lst, Kristiano:2021urj, Kristiano:2022zpn}. These corrections behave nonlinearly, where fluctuations with different wavenumber magnitude can contribute. Therefore, small-scale fluctuations can contribute to the loop corrections of the two-point functions on CMB scales.

As mentioned above, in order to realize PBH formation appropriately, we need an inflation model producing enhanced power spectrum with the amplitude $\mathcal{O}(0.01)$ on a certain small scale while keeping the amplitude at $2.1 \times 10^{-9}$ on CMB scale.  However, such a requirement is only a tree-level statement. So far, understanding of inflation models accommodating PBH formation is very limited beyond tree level, although it has been discussed in 
\cite{Firouzjahi:2018vet, Cheng:2021lif, Syu:2019uwx, Ando:2020fjm, Meng:2022ixx, Chen:2022dah, Inomata:2022yte} for some specific models. 
It is important to ensure that one-loop correction is suppressed compared to the tree-level contribution, so that we can still trust the perturbation theory. 

In our previous papers \cite{Kristiano:2021urj, Kristiano:2022zpn}, we showed that one-loop perturbativity bound can strongly constrain single-field inflation models. We have also qualitatively pointed out a possible problem in PBH formation mechanism. In this Letter, we use a one-loop perturbativity requirement to examine the possibility of PBH formation models in single-field inflation. We calculate contribution of the peak of power spectrum on small scale to one-loop correction of the CMB-scale power spectrum. Requiring one-loop correction to be much smaller than tree-level contribution, we obtain an upper bound on the power spectrum on small scale. 

Specifically, we consider a PBH formation from an extremely flat region in the  potential that leads to a temporary USR motion of the inflaton. At the end, we will explain that our result can be generalized to other PBH formation models in single-field inflation. 

The action of canonical inflation is given by
\begin{equation}
S = \frac{1}{2} \int \md^4x \sqrt{-g} \left[ \mpl^2 R - (\partial_\mu \phi)^2 - 2 V(\phi) \right], \label{action}
\end{equation}
where $\mpl$ is the reduced Planck scale, and $g = \mathrm{det}~g_{\mu\nu}$, $g_{\mu\nu}$ and $R$ are metric tensor and its Ricci scalar. Consider a spatially flat, homogeneous, and isotropic background,
\begin{equation}
\md s^2 = -\md t^2 + a^2(t) \md \mathbf{x}^2 = a^2(\tau) (-\md \tau^2 + \md \mathbf{x}^2),
\end{equation}
where $\tau$ is conformal time. Equations of motion for the scale factor $a(t)$ and the homogeneous part of the inflaton $\phi(t)$ are the Friedmann equations
\begin{equation}
H^2 = \frac{1}{3\mpl^2}\left(\frac{1}{2} \dot{\phi}^2 + V(\phi)\right), ~\dot{H} = - \frac{\dot{\phi}^2}{2 \mpl^2},
\end{equation}
with $H=\dot{a}/{a}$ being the Hubble parameter, and the Klein-Gordon equation
\begin{equation}
\ddot{\phi} + 3 H \dot{\phi} + \frac{\md V}{\md\phi} = 0. \label{klein}
\end{equation}

\begin{figure}
\centering 
\includegraphics[width=\columnwidth]{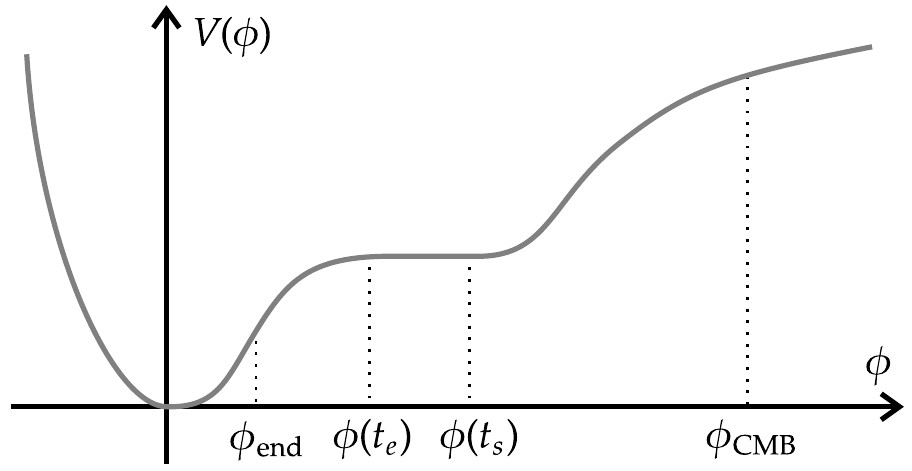}
\caption{\label{fig1} Schematic picture of the inflaton potential realizing PBH formation. When the inflaton is around $\phi_\mathrm{CMB}$, scales probed by CMB observations leave the horizon and it is in the SR regime. It enters an extremely flat region at $t=t_s$  undergoing an USR period. It enters the SR period again at $t=t_e$ until $\phi_\mathrm{end}$, the end of inflation.}
\end{figure}

When CMB-scale fluctuations leave the horizon at around $\phi_\mathrm{CMB}$ (see Fig.~\ref{fig1}), the potential is slightly tilted to realize SR inflation, satisfying 
\begin{equation}
\abs{\frac{\ddot{\phi}}{\dot{\phi} H}} \ll 1, ~\epsilon \equiv - \frac{\dot{H}}{H^2} = \frac{\dot{\phi}^2}{2 \mpl^2 H^2} \ll 1,
\end{equation}
where $\epsilon$ is a SR parameter. In the SR period, $\epsilon$ is approximately constant. Then the inflaton goes through an extremely flat region of the potential, between time $t_s$ to $t_e$, experiencing an USR period. When inflaton enters this region with $\md V / \md \phi \approx 0$, \eqref{klein} becomes $\ddot{\phi} \approx -3H\dot{\phi}$,
so $\dot{\phi} \propto a^{-3}$, which breaks SR approximation \cite{Martin:2012pe}. This makes $\epsilon$ strongly time dependent and extremely small as
\begin{equation}
\epsilon = \frac{\dot{\phi}^2}{2 \mpl^2 H^2}  \propto a^{-6}. \label{epsusr}
\end{equation}
We also define the second SR parameter,
\begin{equation}
\eta \equiv \frac{\dot{\epsilon}}{\epsilon H} = 2 \epsilon +  2 \frac{\ddot{\phi}}{\dot{\phi} H},
\end{equation}
which is approximately constant and very small in the SR period $\abs{\eta} \ll 1$, but large in the USR period $\eta \approx -6$. The latter regime satisfies the condition of the growth of the nonconstant mode of perturbation found in \cite{Saito:2008em}, namely, $3-\epsilon+\eta<0$, so that enhanced spectrum is obtained then.

After the USR period, the inflaton enters the SR period again until the end of inflation. In both the SR and USR period, because $\epsilon$ is very small, the scale factor can be approximated as $a = -1/H\tau \propto e^{H t}$.

Small perturbation from the homogeneous part, $\phi(t)$, of the inflaton $\phi(\mathbf{x}, t)$ and metric can be expressed as 
\begin{gather}
\phi(\mathbf{x},t) = \phi(t) + \delta \phi(\mathbf{x},t), \nonumber\\
\md s^2 = -N^2 \mathrm{d}t^2 + \gamma_{ij} (\mathrm{d}x^i + N^i \mathrm{d}t)(\mathrm{d}x^j + N^j \mathrm{d}t),
\end{gather}
where $\gamma_{ij}$ is the three-dimensional metric on slices of constant $t$, $N$ is the lapse function, and $N^i$ is the shift vector. We choose comoving gauge condition
\begin{equation}
\delta \phi(\mathbf{x},t) = 0, ~\gamma_{ij}(\mathbf{x},t) = a^2(t) e^{2\zeta(\mathbf{x},t)} \delta_{ij},
\end{equation}
where $\zeta(\mathbf{x},t)$ is the curvature perturbation. Here, tensor perturbation is not relevant. Also, $N$ and $N^i$ are obtained by solving constraint equations.

Expanding the action \eqref{action} up to the second-order of the curvature perturbation yields
\begin{equation}
S^{(2)} = M_{\mathrm{pl}}^2 \int \mathrm{d}t ~\mathrm{d}^3x ~a^3 \epsilon  \left[ \dot{\zeta}^2 - \frac{1}{a^2} (\partial_i \zeta)^2  \right].
\label{S2}
\end{equation}
In terms of the Mukhanov-Sasaki variable $v = z \mpl \zeta$, where $z = a \sqrt{2\epsilon}$, the action becomes canonically normalized
\begin{equation}
S^{(2)} = \frac{1}{2} \int \mathrm{d}\tau ~\mathrm{d}^3x \left[ (v')^2 - (\partial_i v)^2 + \frac{z''}{z} v^2 \right],
\end{equation}
where a prime denotes derivative with respect to $\tau$ \cite{Mukhanov:1981xt,Sasaki:1986hm}. In momentum space, quantization is performed by promoting the Mukhanov-Sasaki variable to an operator,
\begin{equation}
v_\bk (\tau) = M_{\mathrm{pl}} z(\tau) \zeta_\bk (\tau)=  v_k(\tau) \hat{a}_\bk + v^*_k (\tau) \hat{a}_{-\bk}^\dagger, \nonumber
\end{equation}
where mode function $v_k(\tau)$ approximately satisfies 
\begin{equation}
v_k'' + \left( k^2 - \frac{2}{\tau^2} \right) v_k = 0
\end{equation}
in both the SR and USR regimes, and the operators satisfy the commutation relation
$[ \hat{a}_{\mathbf{k}}, \hat{a}_{-\mathbf{k'}}^\dagger ] = (2 \pi)^3 \delta(\mathbf{k} + \mathbf{k'})$
under the normalization condition
\begin{equation}
v_k'^* v_k - v_k' v_k^* = i. \label{normalization}
\end{equation}
The general solution of mode function $v_k(\tau)$ is
\begin{equation}
v_k(\tau) = \frac{\mathcal{A}_k}{\sqrt{2k}} \left( 1-\frac{i}{k\tau} \right) e^{-ik\tau} + \frac{\mathcal{B}_k}{\sqrt{2k}} \left( 1+\frac{i}{k\tau} \right) e^{ik\tau},
\end{equation}
where $\mathcal{A}_k$ and $\mathcal{B}_k$ are determined by boundary conditions.
 
At early time, $t \lesssim t_s$, the inflaton was in the SR period with Bunch-Davies initial vacuum, a state $\ket{0}$ defined by $\hat{a}_{\mathbf{k}} \ket{0} = 0$ with $\mathcal{A}_k = 1$ and $\mathcal{B}_k = 0$. Mode function of the curvature perturbation $\zeta_k = v_k/ z \mpl$ is
\begin{equation}
\zeta_k(\tau) = \left( \frac{i H}{2 \mpl \sqrt{\epsilon_{\mathrm{SR}}}} \right)_\star \frac{e^{-ik\tau}}{k^{3/2}}  (1 + ik\tau), \label{zetasr}
\end{equation}
where $\epsilon_{\mathrm{SR}}$ is $\epsilon$ in the SR period and the subscript $\star$ denotes the value at the horizon crossing epoch $\tau = -1/k$. 

At $t_s \lesssim t \lesssim t_e$, the inflaton is in the USR period. We define $\tau_s$ and $\tau_e$ as conformal time corresponding to $t_s$ and $t_e$, respectively. The SR parameter $\epsilon$ can be written as $\epsilon(\tau) = \epsilon_{\mathrm{SR}} (\tau / \tau_s)^6$ from \eqref{epsusr}. Therefore, the curvature perturbation becomes
\begin{align}
\zeta_k(\tau) = & ~\left( \frac{i H}{2 \mpl \sqrt{\epsilon_{\mathrm{SR}}}} \right)_\star  \left( \frac{\tau_s}{\tau} \right)^3 \frac{1}{k^{3/2}} \nonumber\\
& \times\left[ \mathcal{A}_k e^{-ik\tau} (1+ik\tau) - \mathcal{B}_k e^{ik\tau} (1-ik\tau) \right], \label{zetausr}
\end{align}
with $\mathcal{A}_k$ and $\mathcal{B}_k$ determined by matching to the SR solution \eqref{zetasr} at the boundary. We consider instantaneous transition from SR to USR, because it is a good approximation to numerical solutions \cite{Karam:2022nym}. Requiring continuity of $\zeta_k(\tau)$ and $\zeta_k'(\tau)$ at transition $\tau = \tau_s$  \cite{Starobinsky:1992ts, Leach:2001zf, Byrnes:2018txb, Liu:2020oqe, Tasinato:2020vdk, Karam:2022nym, Pi:2022zxs} yields
\begin{equation}
\mathcal{A}_k = 1 - \frac{3(1 + k^2 \tau_s^2)}{2i k^3 \tau_s^3}, ~\mathcal{B}_k = - \frac{3(1 + i k \tau_s)^2}{2i k^3 \tau_s^3} e^{-2ik \tau_s}. \label{coef}
\end{equation}

At late time, $t \gtrsim t_e$, the inflaton goes back to SR dynamics. We define $k_s$ and $k_e$ as wavenumbers crossing the horizon at $\tau_s$ and $\tau_e$, respectively. For perturbation with wavenumber $k \lesssim k_e$, the mode function approaches constant as $\zeta_k (\tau) \approx \zeta_k(\tau_e)$. For perturbation with $k \gtrsim k_e$, the mode function can be approximated as \eqref{zetasr}.

The two-point functions of curvature perturbation and power spectrum at the end of inflation, $\tau_0~(\rightarrow 0)$, can be written as
\begin{gather}
\langle \zeta_\bk(\tau_0) \zeta_{\bk'}(\tau_0) \rangle \equiv (2 \pi)^3 \delta(\bk + \bk') \langle \! \langle \zeta_\bk(\tau_0) \zeta_{-\bk}(\tau_0) \rangle \! \rangle, \\
\Delta^2_s(k) \equiv \frac{k^3}{2 \pi^2} \langle \! \langle \zeta_\bk(\tau_0) \zeta_{-\bk}(\tau_0) \rangle \! \rangle, 
\end{gather}
where the bracket $\langle \cdots \rangle = \bra{0} \cdots \ket{0}$ denotes VEV, and $\Delta^2_s(k)$ is the power spectrum multiplied by the phase space density. For $k \lesssim k_e$, because $\zeta_k (\tau_0) \approx \zeta_k(\tau_e)$, the power spectrum is
\begin{equation}
\Delta^2_{s(0)}(k) = \frac{k^3}{2 \pi^2} \abs{\zeta_k(\tau_0)}^2 \simeq \frac{k^3}{2 \pi^2} \abs{\zeta_k(\tau_e)}^2,
\end{equation}
where $\zeta_k(\tau_e)$ is given by \eqref{zetausr} with coefficients in \eqref{coef}, and the subscript $(0)$ denotes tree-level contribution. 

On a large scale, the power spectrum approaches an almost scale-invariant limit
\begin{equation}
\Delta_{s(\mathrm{SR})}^{2}(k) = \Delta^2_{s(0)}(k \ll k_s) = \left( \frac{H^2}{8 \pi^2 M_{\mathrm{pl}}^2 \epsilon_{\mathrm{SR}}} \right)_\star,
\end{equation}
with a small wavenumber dependence due to the horizon crossing condition manifested in the spectral tilt
\begin{equation}
n_s - 1 = \frac{\md \log \Delta_{s(\mathrm{SR})}^{2}}{\md \log k} = - 2 \epsilon_{\mathrm{SR}} - \eta_{\mathrm{SR}},
\end{equation}
where $\eta_{\mathrm{SR}}$ is $\eta$ in the SR period. This large-scale limit must be consistent with CMB observation. 

\begin{figure}
\centering 
\includegraphics[width=\columnwidth]{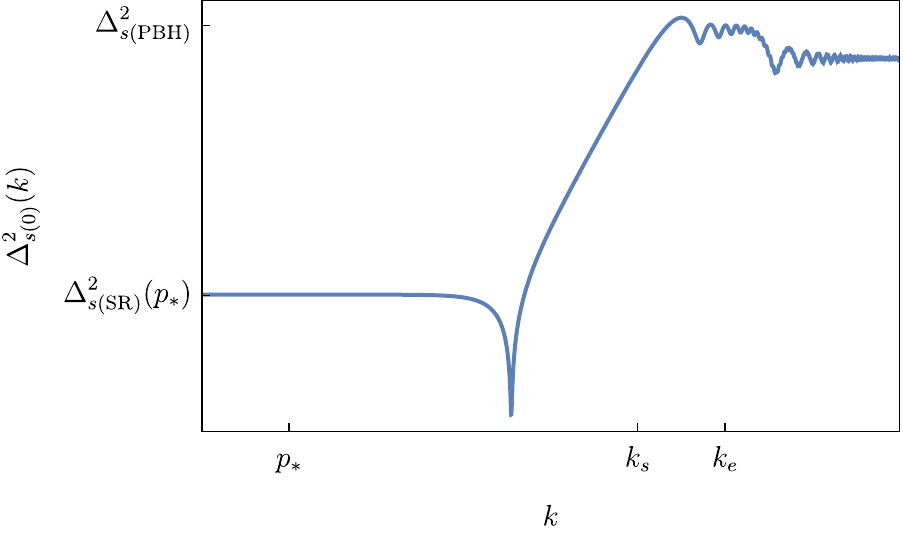}
\caption{\label{fig2} Power spectrum of the curvature perturbation. At CMB scale, $k \ll k_s$, the power spectrum is almost scale invariant. $p_* = 0.05 ~\mathrm{Mpc}^{-1}$ is the pivot scale with amplitude  $\Delta_{s(\mathrm{SR})}^{2}(p_*) = 2.1 \times 10^{-9}$, based on observational result \cite{Planck:2018jri}. At small scale, between $k_s$ and $k_e$, the power spectrum is amplified to typically $\Delta_{s(\mathrm{PBH})}^{2} \sim \mathcal{O}(0.01)$ to form an appreciable amount of PBHs.}
\end{figure}

On a small scale with a larger wavenumber, $k_s \lesssim k \lesssim k_e$, the power spectrum is oscillating. Compared to the large-scale power spectrum, it is enhanced to
\begin{equation}
\Delta_{s(\mathrm{PBH})}^{2} \approx \Delta_{s(\mathrm{SR})}^{2}(k_s) \left( \frac{k_e}{k_s} \right)^6, \label{pspbh}
\end{equation}
whose high density peak may collapse into PBHs. A plot of the typical power spectrum is shown in Fig.~\ref{fig2}.

So far, we have explained the tree-level contribution of the power spectrum. If we expand the action \eqref{action} until higher order in curvature perturbation, we can calculate loop corrections to the power spectrum. Expanding \eqref{action} to third order of $\zeta$ yields the interaction action \cite{Maldacena:2002vr}
\begin{align}
S_{\mathrm{int}} = \mpl^2  \int \md t ~\md^3 x ~a^3 \left[ \epsilon^2  \dot{\zeta}^2 \zeta + \frac{1}{a^2} \epsilon^2 (\partial_i \zeta)^2 \zeta   \right. \nonumber\\
\left. - 2 \epsilon  \dot{\zeta} \partial_i \zeta \partial_i \chi -\frac{1}{2} \epsilon^3 \dot{\zeta}^2 \zeta + \frac{1}{2} \epsilon  \zeta (\partial_i \partial_j \chi)^2 + \frac{1}{2} \epsilon \dot{\eta} \dot{\zeta} \zeta^2 \right], \label{S3}
\end{align}
where $\partial^2 \chi = \epsilon \dot{\zeta}$. In standard SR inflation, the first three terms and the last three terms in \eqref{S3} have coupling $\mathcal{O}(\epsilon^2)$ and $\mathcal{O}(\epsilon^3)$, respectively. The same situation happens in the context of inflation with the PBH formation scenario, except the last term in \eqref{S3}, which has a coupling $\mathcal{O}(\epsilon)$ because $\eta$ can have $\mathcal{O}(1)$ transition \cite{Namjoo:2012aa, Cai:2016ngx, Chen:2013eea, Cai:2018dkf, Davies:2021loj}, approximately from $0$ to $-6$.

We now calculate one-loop correction generated by cubic self-interaction \eqref{S3} in the context of PBH formation using the standard in-in perturbation theory
\begin{align}
\langle \mathcal{O(\tau)} \rangle =  \left\langle \left[ \bar{\mathrm{T}} \exp \left( i \int_{-\infty}^{\tau} \mathrm{d}\tau' H_{\mathrm{int}}(\tau') \right) \right] \mathcal{\hat{O}} (\tau) \right. \nonumber\\
\left. \left[ \mathrm{T} \exp \left( -i \int_{-\infty}^{\tau} \mathrm{d\tau'} H_{\mathrm{int}}(\tau') \right) \right] \right\rangle,
\end{align}
where $\mathcal{\hat{O}}(\tau)$ is an operator at a fixed time $\tau$, and $\mathrm{T}$ and $\bar{\mathrm{T}}$ denote time and antitime ordering. Also, $H_{\mathrm{int}} = -\int \md^3x ~\mathcal{L}_{\mathrm{int}}$ is Hamiltonian corresponding to the Lagrangian $\mathcal{L}_{\mathrm{int}}$, defined by the integrand of \eqref{S3}. In our case, the operator is $\zeta_{\bp}(\tau) \zeta_{-\bp}(\tau) $, where $\mathbf{p}$ is a CMB scale wavevector, evaluated at $\tau=\tau_0 ~(\rightarrow 0)$. 

First-order expansion vanishes, yielding an odd-point correlation function. Second-order expansion of the perturbation theory is
\begin{gather}
\langle \mathcal{O(\tau)} \rangle = \langle \mathcal{O(\tau)} \rangle_{(0,2)}^\dagger + \langle \mathcal{O(\tau)} \rangle_{(1,1)} + \langle \mathcal{O(\tau)} \rangle_{(0,2)}, \label{secin}\\
\langle \mathcal{O(\tau)} \rangle_{(1,1)} = \int_{-\infty}^{\tau} \mathrm{d}\tau_1 \int_{-\infty}^{\tau} \mathrm{d}\tau_2 \langle H_{\mathrm{int}}(\tau_1) \mathcal{\hat{O}} (\tau) H_{\mathrm{int}}(\tau_2) \rangle, \nonumber\\
\langle \mathcal{O(\tau)} \rangle_{(0,2)} = - \int_{-\infty}^{\tau} \mathrm{d}\tau_1 \int_{-\infty}^{\tau_1} \mathrm{d}\tau_2 \langle \mathcal{\hat{O}} (\tau) H_{\mathrm{int}}(\tau_1) H_{\mathrm{int}}(\tau_2) \rangle. \nonumber
\end{gather}

The leading cubic self-interaction is the last term in \eqref{S3} with interaction Hamiltonian \footnote{Strictly speaking, this interaction Hamiltonian is a function of redefined field $\bz$, which was first introduced by Maldacena \cite{Maldacena:2002vr}. The relation between $\zeta$ and $\bz$ is $\zeta = \bz + \frac{1}{4}\eta \bz^2 + \frac{1}{H} \dot{\bz}\bz + \dots$, where dots represent SR or a superhorizon suppressed term. It is shown that the field redefinition method is equivalent to considering boundary interaction \cite{Arroja:2011yj, Burrage:2011hd}. Moreover, quartic self-interaction \cite{Jarnhus:2007ia, Arroja:2008ga} with first-order perturbation theory is expected to generate independent contributions to the one-loop correction because it involves a higher-order SR parameter.}
\begin{equation}
H_{\mathrm{int}} = -\frac{1}{2} \mpl^2 \int \md^3 x ~a^2\epsilon \eta' \zeta' \zeta^2. \label{hint}
\end{equation}
\begin{widetext}
\noindent
After substituting the interaction Hamiltonian to the perturbation theory, we find 
\begin{gather}
\langle \zeta_{\bp}(\tau_0) \zeta_{-\bp}(\tau_0) \rangle_{(1,1)} = \frac{1}{4} \mpl^4 \int_{-\infty}^{\tau_0} \md \tau_1 ~a^2(\tau_1) \epsilon(\tau_1) \eta'(\tau_1) \int_{-\infty}^{\tau_0} \md \tau_2 ~a^2(\tau_2) \epsilon(\tau_2) \eta'(\tau_2)  \nonumber \\
\times \int \prod_{a = 1}^6 \left[ \frac{\md^3 k_a}{(2\pi)^3} \right] \delta(\bk_1+\bk_2+\bk_3) \delta(\bk_4+\bk_5+\bk_6) \left\langle \zeta_{\bk_1}'(\tau_1) \zeta_{\bk_2}(\tau_1) \zeta_{\bk_3}(\tau_1) \zeta_{\bp}(\tau_0) \zeta_{-\bp}(\tau_0) \zeta_{\bk_4}'(\tau_2) \zeta_{\bk_5}(\tau_2) \zeta_{\bk_6}(\tau_2) \right\rangle, \label{onel1} \\
\langle \zeta_{\bp}(\tau_0) \zeta_{-\bp}(\tau_0) \rangle_{(0,2)} = -\frac{1}{4} \mpl^4 \int_{-\infty}^{\tau_0} \md \tau_1 ~a^2(\tau_1) \epsilon(\tau_1) \eta'(\tau_1) \int_{-\infty}^{\tau_1} \md \tau_2 ~a^2(\tau_2) \epsilon(\tau_2) \eta'(\tau_2) \nonumber \\
\times \int \prod_{a = 1}^6 \left[ \frac{\md^3 k_a}{(2\pi)^3} \right] \delta(\bk_1+\bk_2+\bk_3) \delta(\bk_4+\bk_5+\bk_6) \left\langle \zeta_{\bp}(\tau_0) \zeta_{-\bp}(\tau_0) \zeta_{\bk_1}'(\tau_1) \zeta_{\bk_2}(\tau_1) \zeta_{\bk_3}(\tau_1) \zeta_{\bk_4}'(\tau_2) \zeta_{\bk_5}(\tau_2) \zeta_{\bk_6}(\tau_2) \right\rangle. \label{onel2}
\end{gather}
To evaluate the time integral, we note that $\eta$ is almost constant in both the SR and USR periods, so $\eta'(\tau) \approx 0$ except for sharp transitions around $\tau = \tau_s$ and $\tau = \tau_e$. Therefore, the time integral can be evaluated as
\begin{equation}
 \int_{-\infty}^0 \md\tau ~\eta'(\tau) f(\tau) = \Delta\eta ~f(\tau_e), \label{tint}
\end{equation}
where $f(\tau)$ is a general continuous function. We neglect contribution from $\tau = \tau_s$ because it is much smaller than that from $\tau = \tau_e$. Nevertheless, contribution from $\tau = \tau_s$ is essential to recover Maldacena's theorem for the bispectrum, as we show in our companion paper \cite{Kristiano:2023scm}. Performing operator expansion and Wick contraction, we obtain total one-loop correction
\begin{align}
& \langle \! \langle \zeta_{\bp}(\tau_0) \zeta_{-\bp}(\tau_0) \rangle \! \rangle_{(1)} = \langle \! \langle \zeta_{\bp}(\tau_0) \zeta_{-\bp}(\tau_0) \rangle \! \rangle_{(1,1)} + 2 \mathrm{Re} \langle \! \langle \zeta_{\bp}(\tau_0) \zeta_{-\bp}(\tau_0) \rangle \! \rangle_{(0,2)} = \frac{1}{4} \mpl^4 \epsilon^2(\tau_e) a^4(\tau_e) (\Delta\eta)^2 \\& 
\times \int \frac{\md^3 k}{(2\pi)^3} \left\lbrace \zeta_p(\tau_0) \zeta_p^*(\tau_0) \left( 4 \zeta_p' \zeta_p'^* \zeta_q \zeta_q^* \zeta_k \zeta_k^* 
+ 8 \zeta_p'^* \zeta_p \zeta_q' \zeta_q^* \zeta_k \zeta_k^*  + 8 \zeta_p' \zeta_p^* \zeta_q'^* \zeta_q \zeta_k \zeta_k^* + 16 \zeta_p \zeta_p^* \zeta_q' \zeta_q'^* \zeta_k \zeta_k^* \right)_{\tau = \tau_e} \right.  \nonumber\\ & \left.  
- \mathrm{Re} \left[ \zeta_p(\tau_0)\zeta_p(\tau_0) \left( 4  \zeta_p'^* \zeta_p'^* \zeta_q \zeta_q^* \zeta_k \zeta_k^* + 8  \zeta_p'^* \zeta_p^* \zeta_q' \zeta_q^* \zeta_k \zeta_k^* + 8  \zeta_p'^* \zeta_p^* \zeta_q'^* \zeta_q \zeta_k \zeta_k^* + 16 \zeta_p^* \zeta_p^* \zeta_q' \zeta_q'^* \zeta_k \zeta_k^* \right)_{\tau = \tau_e} \right] \right\rbrace, \nonumber
\end{align}
where $\mathbf{q} = \mathbf{k} - \mathbf{p}$. After some algebra, for $p\ll k$, the leading term is simplified to
\begin{equation}
\langle \! \langle \zeta_{\bp}(\tau_0) \zeta_{-\bp}(\tau_0) \rangle \! \rangle_{(1)} = \frac{1}{4} \mpl^4 \epsilon^2(\tau_e) a^4(\tau_e) (\Delta\eta)^2 \abs{\zeta_p(\tau_0)}^2 16\int \frac{\md^3 k}{(2\pi)^3} \left[ \abs{\zeta_k}^2 ~\mathrm{Im}(\zeta_p \zeta_p'^*) ~\mathrm{Im}(\zeta_q \zeta_q'^*) \right]_{\tau = \tau_e}. \label{loop}
\end{equation}
\end{widetext}
Expressing \eqref{normalization} in terms of curvature perturbation during the USR period, we obtain
\begin{equation}
[\mathrm{Im}(\zeta_q \zeta_q'^*)]_{\tau = \tau_e} = \frac{1}{4 \mpl^2 \epsilon(\tau_e) a^2(\tau_e) },
\end{equation}
which is independent of wavenumber $q$. It leads to the commutation relation $[\zeta, \dot{\zeta}] \propto a^3$, which is in line with large quantum loop correction that we will find shortly.
Then, substituting it to \eqref{loop} yields one-loop correction to the power spectrum
\begin{equation}
\Delta_{s(1)}^2(p) = \frac{1}{4} (\Delta\eta)^2  \Delta_{s(0)}^{2}(p)  \int_{k_{\mathrm{IR}}}^{k_{\mathrm{UV}}} \frac{\md k}{2 \pi^2} k^2 \abs{\zeta_k(\tau_e)}^2. \label{loop2}
\end{equation}

In principle, the wavenumber integral should be extended from well below CMB pivot scale $k_{\mathrm{IR}} \ll p_*= 0.05 ~\mathrm{Mpc}^{-1}$ to the ultraviolet (UV) cutoff scale $k_{\mathrm{UV}} \rightarrow \infty$. Contributions from a large length scale including the CMB scale can be neglected because they are much smaller than those from amplified perturbation on a small scale. On the other hand, contribution from the UV scale diverges. Because we are interested in the finite effect of the  amplified perturbation on a specific small scale due to the USR period to one-loop correction, we conservatively restrict the wavenumber integration domain from $k_{\mathrm{IR}}=k_s$ to $k_{\mathrm{UV}}=k_e$. The UV cutoff issue will be discussed later. For $k_{\mathrm{UV}}/k_{\mathrm{IR}}=  k_e/k_s \gg 1$, \eqref{loop2} reads as 
\begin{equation}
\Delta_{s(1)}^2(p) = \frac{1}{4} (\Delta\eta)^2  \Delta_{s(0)}^{2}(p) \Delta_{s(\mathrm{PBH})}^{2} \left( 1.1 + \log\frac{k_e}{k_s} \right).
\end{equation}

In order for the standard cosmological perturbation theory to be trustworthy, one-loop correction must be much smaller than the tree-level contribution, namely $\Delta_{s(1)}^2 \ll \Delta_{s(0)}^{2}$. It leads to a strong inequality,
\begin{equation}
\frac{1}{4} (\Delta\eta)^2 \Delta_{s(\mathrm{PBH})}^{2} \left( 1.1 + \log\frac{k_e}{k_s} \right) \ll 1. \label{ineq}
\end{equation}
We can obtain a bound on $k_e/k_s$ by substituting numerical values $\Delta_{s(\mathrm{SR})}^{2}(p_*) = 2.1 \times 10^{-9}$ and $n_s=0.9649 \pm 0.0042$ at pivot scale, based on observational result \cite{Planck:2018jri}. Also, $\Delta\eta = 6$ as a transition from the USR to SR period. Solving inequality \eqref{ineq} numerically leads to an upper bound $k_e/k_s \ll 15$, which is equivalent to
\begin{equation}
\Delta_{s(\mathrm{PBH})}^{2} \ll \frac{1}{(\Delta\eta)^2} \simeq 0.03 . \label{ineq2}
\end{equation}

So far, we have considered the finite contribution of the superhorizon perturbation at $\tau = \tau_e$ to the one-loop correction. In addition, there is a contribution from perturbations which are still inside the horizon at  $\tau = \tau_e$. To identify such contributions, we introduce a physical UV cutoff $\Lambda$ to the upper bound of the wavenumber integral such as \footnote{This can be derived by introducing physical wavenumber cutoff $\Lambda a(\tau)$ to the upper bound of the wavenumber integral in \eqref{onel1} and \eqref{onel2} \cite{Senatore:2009cf}. After performing a time integral \eqref{tint}, the cutoff is evaluated at $\tau = \tau_e$, which leads to \eqref{intlambda}}
\begin{equation}
\int_{k_{\mathrm{IR}}}^{k_{\mathrm{UV}}} \frac{\md k}{k} \Delta^2_{s(0)}(k) = \left( \int_{k_s}^{k_e} + \int_{k_e}^{\Lambda a(\tau_e)} \right) \frac{\md k}{k} \Delta^2_{s(0)}(k). \label{intlambda}
\end{equation}
Including the divergence, the one-loop correction reads
\begin{align}
\Delta_{s(1)}^2(p) = \frac{1}{4} (\Delta\eta)^2  \Delta_{s(0)}^{2}(p) \Delta_{s(\mathrm{PBH})}^{2}  \nonumber\\
\times  \left( 1.1 + \log\frac{k_e}{k_s} + \log\frac{\Lambda}{H} + \frac{\Lambda^2 - H^2}{2H^2} \right).
\end{align}

The bare power spectrum at one-loop order is given by
\begin{gather}
\Delta_s^2(p) = \Delta_{s(0)}^{2}(p_*) \left( \frac{p}{p_*} \right)^{n_s-1} \left\lbrace 1 + \frac{1}{4} (\Delta\eta)^2 \Delta_{s(\mathrm{PBH})}^{2} \right. \nonumber\\
\left. \times \left( 1.1 + \log\frac{k_e}{k_s} + \log\tilde{\Lambda} + \frac{\tilde{\Lambda}^2 - 1}{2} \right)  \right. \nonumber\\
\left. + \mathcal{O}[ ( (\Delta\eta)^2  \Delta_{s(\mathrm{PBH})}^{2} )^2 ] \right\rbrace,
\end{gather}
where $\tilde{\Lambda} = \Lambda / H$ is defined. Such a power spectrum diverges as we take the $\Lambda \rightarrow \infty$ limit \footnote{Dimensional regularization leads to a similar expression, except $\log \tilde{\Lambda}$ is changed to $1/\delta$, where $\delta$ is a small correction to the spatial dimension. Another difference is that dimensional regularization cannot capture the quadratic divergence $\tilde{\Lambda}^2$, as expected. Compared to one-loop computation for the case considered in \cite{Senatore:2009cf, delRio:2018vrj}, the time integral in our case is dominated by $\tau = \tau_e$, not the CMB horizon crossing time $\tau = -1/p$.}. To make it finite, we need to renormalize order by order the coefficient $\Delta_{s(0)}^{2}(p_*)$ so it depends on $\tilde{\Lambda}$. This procedure is equivalent to introducing a counterterm $\delta Z$, defined in $\Delta_{s(0)}^{2}(p_*) \equiv (1+ \delta Z) \Delta_{s(0)}^{2}(p_*, \tilde{\mu}) $. At one-loop order, after substituting \eqref{pspbh}, $\delta Z$ can be read off from the renormalized coefficient
\begin{gather}
\Delta_{s(0)}^2(p_*) \equiv \Delta_{s(0)}^{2}(p_*, \tilde{\mu}) \left\lbrace 1 + \frac{1}{4} (\Delta\eta)^2 \Delta_{s(0)}^{2}(p_*, \tilde{\mu}) \right. \nonumber\\
\left. \times \left( \frac{k_e}{k_s} \right)^6 \left( -1.1 - \log\frac{k_e}{k_s} + \log\frac{\tilde{\mu}}{\tilde{\Lambda}} + \frac{\tilde{\mu}^2 - \tilde{\Lambda}^2}{2} \right) \right. \nonumber\\
\left. + \mathcal{O}[ ( (\Delta\eta)^2  \Delta_{s(\mathrm{PBH})}^{2} )^2 ] \right\rbrace,
\end{gather}
where $\tilde{\mu} = \mu / H$ with $\mu$ being a renormalization scale. Therefore, we obtained the renormalized power spectrum
\begin{gather}
\Delta_s^2(p) = \Delta_{s(0)}^{2}(p_*, \tilde{\mu})  \left( \frac{p}{p_*} \right)^{n_s-1}    \nonumber\\
\times \left\lbrace 1 + \frac{1}{4} (\Delta\eta)^2 \Delta_{s(0)}^{2}(p_*, \tilde{\mu})\left( \frac{k_e}{k_s} \right)^6 \left( \log\tilde{\mu} + \frac{\tilde{\mu}^2 - 1}{2} \right) \right.  \nonumber\\
\left. + \mathcal{O}[ ( (\Delta\eta)^2  \Delta_{s(\mathrm{PBH})}^{2} )^2 ] \right\rbrace. 
\end{gather}
Taking renormalization scale $\mu = H$, the renormalized power spectrum reads as
\begin{align}
\Delta_s^2(p) = & ~\Delta_{s(0)}^{2}(p_*, \mu = H) \left( \frac{p}{p_*} \right)^{n_s-1} \nonumber\\
& \times \left\lbrace 1 + \mathcal{O}[ ( (\Delta\eta)^2  \Delta_{s(\mathrm{PBH})}^{2} )^2 ] \right\rbrace,
\end{align}
where $\Delta_{s(0)}^{2}(p_*, \mu = H)$ is fixed to the observed power spectrum $2.1 \times 10^{-9}$.

Such order by order renormalization is justified if higher-order loop correction is suppressed compared to the lower-order one. In this context, this is possible only if $(\Delta\eta)^2  \Delta_{s(\mathrm{PBH})}^{2} \ll 1$, which is the same condition as \eqref{ineq2}. Therefore, condition \eqref{ineq2} is a necessary condition for the theory to be perturbative, or a requirement to avoid strong coupling. Indeed, this perturbative coupling requirement is needed to justify the use of in-in perturbation theory \eqref{secin}. Although we focus on loop correction to large-scale power spectrum, such coupling also affect the small-scale power spectrum and the same constraint is needed to avoid strong coupling. 

In single-field inflation, PBH formation models can be classified into two categories \cite{Karam:2022nym}. The first category is models with features in the inflationary potential or nonminimally coupled inflaton with potential defined in the Einstein frame. Models with an extremely flat feature fall in this category including those with an inflection point in the potential \cite{Ivanov:1994pa, Kawasaki:1997ju, Kawasaki:1998vx, Kawasaki:2016pql, Gao:2018pvq, Nanopoulos:2020nnh, Wu:2021zta, Stamou:2021qdk, Cheng:2018yyr, Ballesteros:2019hus, Pi:2017gih, Cicoli:2018asa, Ozsoy:2018flq, Cicoli:2022sih, Ezquiaga:2017fvi, Ballesteros:2017fsr, Drees:2019xpp, Cheong:2019vzl, Rasanen:2018fom, Dalianis:2019asr, Dalianis:2018frf, Mahbub:2019uhl, Garcia-Bellido:2017mdw, Ragavendra:2020sop}. Other examples of features are a tiny bump or dip \cite{Mishra:2019pzq, Atal:2019cdz, Zheng:2021vda, Wang:2021kbh, Rezazadeh:2021clf, Iacconi:2021ltm}, an upward or downward step \cite{Cai:2021zsp, Kefala:2020xsx, Ng:2021hll, Motohashi:2019rhu, Inomata:2021uqj, Inomata:2021tpx}, polynomial shape \cite{Hertzberg:2017dkh, Ballesteros:2020qam, Kannike:2017bxn, Di:2017ndc}, and Coleman-Weinberg potential \cite{Yokoyama:1998pt, Saito:2008em, Bugaev:2008bi}. In these examples, modification of potential makes a sharp $\eta$ transition on the inflaton dynamics. 

The second category is models with modified gravity or beyond nonminimally coupled inflaton, for example, models  based on $k$ \cite{Armendariz-Picon:1999hyi} or $G$ \cite{Kobayashi:2011nu} inflation \cite{Solbi:2021wbo, Solbi:2021rse, Teimoori:2021pte}, the effective field theory of inflation \cite{Ballesteros:2018wlw, Ballesteros:2021fsp, Ashoorioon:2019xqc}, $f(R)$ gravity \cite{Frolovsky:2022ewg}, a nonminimal derivative coupling \cite{Fu:2019ttf, Heydari:2021gea}, Gauss-Bonnet inflation \cite{Kawai:2021edk}, and bumpy axion inflation \cite{Ozsoy:2020kat}. In these examples, amplification of small-scale perturbation can be realized by a sharp transition of $\eta$ and/or other parameters. In \cite{Ballesteros:2018wlw, Ballesteros:2021fsp}, amplification of small-scale perturbation is caused by a sharp transition of the sound speed, $c_s$, a quantity that parametrizes deviation from a canonical kinetic term. In this case, coupling $\eta'$ in cubic self-interaction \eqref{hint} is modified to $(\eta / c_s^2)'$ \cite{Chen:2006nt, Seery:2005wm}. For this category, one-loop correction should be investigated further.

In conclusion, we have calculated the one-loop correction of the inflationary power spectrum in single-field inflation realizing PBH formation. We have shown that models realizing an appreciable amount of PBH formation with the enhanced small-scale spectrum by USR inflaton dynamics inevitably induces nonperturbative coupling to the power spectrum on a CMB scale. Requiring such a coupling to be perturbative so that cosmological perturbation theory is valid on large scales, we therefore conclude a strict upper bound on the small-scale power spectrum associated with transition of the second SR parameter.

J.~K. thanks Hayato Motohashi for a fruitful discussion. J.~K. acknowledges the support from JSPS KAKENHI Grants No.~22KJ1006 and No.~22J20289. J.~K. is also supported by the Global Science Graduate Course (GSGC) program of The University of Tokyo. J.~Y. is supported by JSPS KAKENHI Grant No.~20H05639 and Grant-in-Aid for Scientific Research on Innovative Areas 20H05248.

\bibliographystyle{apsrev4-1}
\bibliography{reference}

\providecommand{\noopsort}[1]{}\providecommand{\singleletter}[1]{#1}%
\begin{thebibliography}{152}%
\makeatletter
\providecommand \@ifxundefined [1]{%
 \@ifx{#1\undefined}
}%
\providecommand \@ifnum [1]{%
 \ifnum #1\expandafter \@firstoftwo
 \else \expandafter \@secondoftwo
 \fi
}%
\providecommand \@ifx [1]{%
 \ifx #1\expandafter \@firstoftwo
 \else \expandafter \@secondoftwo
 \fi
}%
\providecommand \natexlab [1]{#1}%
\providecommand \enquote  [1]{``#1''}%
\providecommand \bibnamefont  [1]{#1}%
\providecommand \bibfnamefont [1]{#1}%
\providecommand \citenamefont [1]{#1}%
\providecommand \href@noop [0]{\@secondoftwo}%
\providecommand \href [0]{\begingroup \@sanitize@url \@href}%
\providecommand \@href[1]{\@@startlink{#1}\@@href}%
\providecommand \@@href[1]{\endgroup#1\@@endlink}%
\providecommand \@sanitize@url [0]{\catcode `\\12\catcode `\$12\catcode `\&12\catcode `\#12\catcode `\^12\catcode `\_12\catcode `\%12\relax}%
\providecommand \@@startlink[1]{}%
\providecommand \@@endlink[0]{}%
\providecommand \url  [0]{\begingroup\@sanitize@url \@url }%
\providecommand \@url [1]{\endgroup\@href {#1}{\urlprefix }}%
\providecommand \urlprefix  [0]{URL }%
\providecommand \Eprint [0]{\href }%
\providecommand \doibase [0]{http://dx.doi.org/}%
\providecommand \selectlanguage [0]{\@gobble}%
\providecommand \bibinfo  [0]{\@secondoftwo}%
\providecommand \bibfield  [0]{\@secondoftwo}%
\providecommand \translation [1]{[#1]}%
\providecommand \BibitemOpen [0]{}%
\providecommand \bibitemStop [0]{}%
\providecommand \bibitemNoStop [0]{.\EOS\space}%
\providecommand \EOS [0]{\spacefactor3000\relax}%
\providecommand \BibitemShut  [1]{\csname bibitem#1\endcsname}%
\let\auto@bib@innerbib\@empty
\bibitem [{\citenamefont {{Zel'dovich}}\ and\ \citenamefont {{Novikov}}(1967)}]{Zel:1967}%
  \BibitemOpen
  \bibfield  {author} {\bibinfo {author} {\bibfnamefont {Y.~B.}\ \bibnamefont {{Zel'dovich}}}\ and\ \bibinfo {author} {\bibfnamefont {I.~D.}\ \bibnamefont {{Novikov}}},\ }\href@noop {} {\bibfield  {journal} {\bibinfo  {journal} {Sov. Astron.}\ }\textbf {\bibinfo {volume} {10}},\ \bibinfo {pages} {602} (\bibinfo {year} {1967})}\BibitemShut {NoStop}%
\bibitem [{\citenamefont {Hawking}(1971)}]{Hawking:1971ei}%
  \BibitemOpen
  \bibfield  {author} {\bibinfo {author} {\bibfnamefont {S.}~\bibnamefont {Hawking}},\ }\href@noop {} {\bibfield  {journal} {\bibinfo  {journal} {Mon. Not. Roy. Astron. Soc.}\ }\textbf {\bibinfo {volume} {152}},\ \bibinfo {pages} {75} (\bibinfo {year} {1971})}\BibitemShut {NoStop}%
\bibitem [{\citenamefont {Carr}\ and\ \citenamefont {Hawking}(1974)}]{Carr:1974nx}%
  \BibitemOpen
  \bibfield  {author} {\bibinfo {author} {\bibfnamefont {B.~J.}\ \bibnamefont {Carr}}\ and\ \bibinfo {author} {\bibfnamefont {S.~W.}\ \bibnamefont {Hawking}},\ }\href@noop {} {\bibfield  {journal} {\bibinfo  {journal} {Mon. Not. Roy. Astron. Soc.}\ }\textbf {\bibinfo {volume} {168}},\ \bibinfo {pages} {399} (\bibinfo {year} {1974})}\BibitemShut {NoStop}%
\bibitem [{\citenamefont {Hawking}(1974)}]{Hawking:1974rv}%
  \BibitemOpen
  \bibfield  {author} {\bibinfo {author} {\bibfnamefont {S.~W.}\ \bibnamefont {Hawking}},\ }\href {\doibase 10.1038/248030a0} {\bibfield  {journal} {\bibinfo  {journal} {Nature}\ }\textbf {\bibinfo {volume} {248}},\ \bibinfo {pages} {30} (\bibinfo {year} {1974})}\BibitemShut {NoStop}%
\bibitem [{\citenamefont {Chapline}(1975)}]{Chapline:1975ojl}%
  \BibitemOpen
  \bibfield  {author} {\bibinfo {author} {\bibfnamefont {G.~F.}\ \bibnamefont {Chapline}},\ }\href {\doibase 10.1038/253251a0} {\bibfield  {journal} {\bibinfo  {journal} {Nature}\ }\textbf {\bibinfo {volume} {253}},\ \bibinfo {pages} {251} (\bibinfo {year} {1975})}\BibitemShut {NoStop}%
\bibitem [{\citenamefont {Garcia-Bellido}\ \emph {et~al.}(1996)\citenamefont {Garcia-Bellido}, \citenamefont {Linde},\ and\ \citenamefont {Wands}}]{Garcia-Bellido:1996mdl}%
  \BibitemOpen
  \bibfield  {author} {\bibinfo {author} {\bibfnamefont {J.}~\bibnamefont {Garcia-Bellido}}, \bibinfo {author} {\bibfnamefont {A.~D.}\ \bibnamefont {Linde}}, \ and\ \bibinfo {author} {\bibfnamefont {D.}~\bibnamefont {Wands}},\ }\href {\doibase 10.1103/PhysRevD.54.6040} {\bibfield  {journal} {\bibinfo  {journal} {Phys. Rev. D}\ }\textbf {\bibinfo {volume} {54}},\ \bibinfo {pages} {6040} (\bibinfo {year} {1996})},\ \Eprint {http://arxiv.org/abs/astro-ph/9605094} {arXiv:astro-ph/9605094} \BibitemShut {NoStop}%
\bibitem [{\citenamefont {Ivanov}\ \emph {et~al.}(1994)\citenamefont {Ivanov}, \citenamefont {Naselsky},\ and\ \citenamefont {Novikov}}]{Ivanov:1994pa}%
  \BibitemOpen
  \bibfield  {author} {\bibinfo {author} {\bibfnamefont {P.}~\bibnamefont {Ivanov}}, \bibinfo {author} {\bibfnamefont {P.}~\bibnamefont {Naselsky}}, \ and\ \bibinfo {author} {\bibfnamefont {I.}~\bibnamefont {Novikov}},\ }\href {\doibase 10.1103/PhysRevD.50.7173} {\bibfield  {journal} {\bibinfo  {journal} {Phys. Rev. D}\ }\textbf {\bibinfo {volume} {50}},\ \bibinfo {pages} {7173} (\bibinfo {year} {1994})}\BibitemShut {NoStop}%
\bibitem [{\citenamefont {Yokoyama}(1997)}]{Yokoyama:1995ex}%
  \BibitemOpen
  \bibfield  {author} {\bibinfo {author} {\bibfnamefont {J.}~\bibnamefont {Yokoyama}},\ }\href@noop {} {\bibfield  {journal} {\bibinfo  {journal} {Astron. Astrophys.}\ }\textbf {\bibinfo {volume} {318}},\ \bibinfo {pages} {673} (\bibinfo {year} {1997})},\ \Eprint {http://arxiv.org/abs/astro-ph/9509027} {arXiv:astro-ph/9509027} \BibitemShut {NoStop}%
\bibitem [{\citenamefont {Afshordi}\ \emph {et~al.}(2003)\citenamefont {Afshordi}, \citenamefont {McDonald},\ and\ \citenamefont {Spergel}}]{Afshordi:2003zb}%
  \BibitemOpen
  \bibfield  {author} {\bibinfo {author} {\bibfnamefont {N.}~\bibnamefont {Afshordi}}, \bibinfo {author} {\bibfnamefont {P.}~\bibnamefont {McDonald}}, \ and\ \bibinfo {author} {\bibfnamefont {D.~N.}\ \bibnamefont {Spergel}},\ }\href {\doibase 10.1086/378763} {\bibfield  {journal} {\bibinfo  {journal} {Astrophys. J. Lett.}\ }\textbf {\bibinfo {volume} {594}},\ \bibinfo {pages} {L71} (\bibinfo {year} {2003})},\ \Eprint {http://arxiv.org/abs/astro-ph/0302035} {arXiv:astro-ph/0302035} \BibitemShut {NoStop}%
\bibitem [{\citenamefont {Frampton}\ \emph {et~al.}(2010)\citenamefont {Frampton}, \citenamefont {Kawasaki}, \citenamefont {Takahashi},\ and\ \citenamefont {Yanagida}}]{Frampton:2010sw}%
  \BibitemOpen
  \bibfield  {author} {\bibinfo {author} {\bibfnamefont {P.~H.}\ \bibnamefont {Frampton}}, \bibinfo {author} {\bibfnamefont {M.}~\bibnamefont {Kawasaki}}, \bibinfo {author} {\bibfnamefont {F.}~\bibnamefont {Takahashi}}, \ and\ \bibinfo {author} {\bibfnamefont {T.~T.}\ \bibnamefont {Yanagida}},\ }\href {\doibase 10.1088/1475-7516/2010/04/023} {\bibfield  {journal} {\bibinfo  {journal} {JCAP}\ }\textbf {\bibinfo {volume} {04}},\ \bibinfo {pages} {023} (\bibinfo {year} {2010})},\ \Eprint {http://arxiv.org/abs/1001.2308} {arXiv:1001.2308 [hep-ph]} \BibitemShut {NoStop}%
\bibitem [{\citenamefont {Belotsky}\ \emph {et~al.}(2014)\citenamefont {Belotsky}, \citenamefont {Dmitriev}, \citenamefont {Esipova}, \citenamefont {Gani}, \citenamefont {Grobov}, \citenamefont {Khlopov}, \citenamefont {Kirillov}, \citenamefont {Rubin},\ and\ \citenamefont {Svadkovsky}}]{Belotsky:2014kca}%
  \BibitemOpen
  \bibfield  {author} {\bibinfo {author} {\bibfnamefont {K.~M.}\ \bibnamefont {Belotsky}}, \bibinfo {author} {\bibfnamefont {A.~D.}\ \bibnamefont {Dmitriev}}, \bibinfo {author} {\bibfnamefont {E.~A.}\ \bibnamefont {Esipova}}, \bibinfo {author} {\bibfnamefont {V.~A.}\ \bibnamefont {Gani}}, \bibinfo {author} {\bibfnamefont {A.~V.}\ \bibnamefont {Grobov}}, \bibinfo {author} {\bibfnamefont {M.~Y.}\ \bibnamefont {Khlopov}}, \bibinfo {author} {\bibfnamefont {A.~A.}\ \bibnamefont {Kirillov}}, \bibinfo {author} {\bibfnamefont {S.~G.}\ \bibnamefont {Rubin}}, \ and\ \bibinfo {author} {\bibfnamefont {I.~V.}\ \bibnamefont {Svadkovsky}},\ }\href {\doibase 10.1142/S0217732314400057} {\bibfield  {journal} {\bibinfo  {journal} {Mod. Phys. Lett. A}\ }\textbf {\bibinfo {volume} {29}},\ \bibinfo {pages} {1440005} (\bibinfo {year} {2014})},\ \Eprint {http://arxiv.org/abs/1410.0203} {arXiv:1410.0203 [astro-ph.CO]} \BibitemShut {NoStop}%
\bibitem [{\citenamefont {Carr}\ \emph {et~al.}(2016)\citenamefont {Carr}, \citenamefont {Kuhnel},\ and\ \citenamefont {Sandstad}}]{Carr:2016drx}%
  \BibitemOpen
  \bibfield  {author} {\bibinfo {author} {\bibfnamefont {B.}~\bibnamefont {Carr}}, \bibinfo {author} {\bibfnamefont {F.}~\bibnamefont {Kuhnel}}, \ and\ \bibinfo {author} {\bibfnamefont {M.}~\bibnamefont {Sandstad}},\ }\href {\doibase 10.1103/PhysRevD.94.083504} {\bibfield  {journal} {\bibinfo  {journal} {Phys. Rev. D}\ }\textbf {\bibinfo {volume} {94}},\ \bibinfo {pages} {083504} (\bibinfo {year} {2016})},\ \Eprint {http://arxiv.org/abs/1607.06077} {arXiv:1607.06077 [astro-ph.CO]} \BibitemShut {NoStop}%
\bibitem [{\citenamefont {Inomata}\ \emph {et~al.}(2017)\citenamefont {Inomata}, \citenamefont {Kawasaki}, \citenamefont {Mukaida}, \citenamefont {Tada},\ and\ \citenamefont {Yanagida}}]{Inomata:2017okj}%
  \BibitemOpen
  \bibfield  {author} {\bibinfo {author} {\bibfnamefont {K.}~\bibnamefont {Inomata}}, \bibinfo {author} {\bibfnamefont {M.}~\bibnamefont {Kawasaki}}, \bibinfo {author} {\bibfnamefont {K.}~\bibnamefont {Mukaida}}, \bibinfo {author} {\bibfnamefont {Y.}~\bibnamefont {Tada}}, \ and\ \bibinfo {author} {\bibfnamefont {T.~T.}\ \bibnamefont {Yanagida}},\ }\href {\doibase 10.1103/PhysRevD.96.043504} {\bibfield  {journal} {\bibinfo  {journal} {Phys. Rev. D}\ }\textbf {\bibinfo {volume} {96}},\ \bibinfo {pages} {043504} (\bibinfo {year} {2017})},\ \Eprint {http://arxiv.org/abs/1701.02544} {arXiv:1701.02544 [astro-ph.CO]} \BibitemShut {NoStop}%
\bibitem [{\citenamefont {Espinosa}\ \emph {et~al.}(2018)\citenamefont {Espinosa}, \citenamefont {Racco},\ and\ \citenamefont {Riotto}}]{Espinosa:2017sgp}%
  \BibitemOpen
  \bibfield  {author} {\bibinfo {author} {\bibfnamefont {J.~R.}\ \bibnamefont {Espinosa}}, \bibinfo {author} {\bibfnamefont {D.}~\bibnamefont {Racco}}, \ and\ \bibinfo {author} {\bibfnamefont {A.}~\bibnamefont {Riotto}},\ }\href {\doibase 10.1103/PhysRevLett.120.121301} {\bibfield  {journal} {\bibinfo  {journal} {Phys. Rev. Lett.}\ }\textbf {\bibinfo {volume} {120}},\ \bibinfo {pages} {121301} (\bibinfo {year} {2018})},\ \Eprint {http://arxiv.org/abs/1710.11196} {arXiv:1710.11196 [hep-ph]} \BibitemShut {NoStop}%
\bibitem [{\citenamefont {Green}\ and\ \citenamefont {Kavanagh}(2021)}]{Green:2020jor}%
  \BibitemOpen
  \bibfield  {author} {\bibinfo {author} {\bibfnamefont {A.~M.}\ \bibnamefont {Green}}\ and\ \bibinfo {author} {\bibfnamefont {B.~J.}\ \bibnamefont {Kavanagh}},\ }\href {\doibase 10.1088/1361-6471/abc534} {\bibfield  {journal} {\bibinfo  {journal} {J. Phys. G}\ }\textbf {\bibinfo {volume} {48}},\ \bibinfo {pages} {043001} (\bibinfo {year} {2021})},\ \Eprint {http://arxiv.org/abs/2007.10722} {arXiv:2007.10722 [astro-ph.CO]} \BibitemShut {NoStop}%
\bibitem [{\citenamefont {Carr}\ and\ \citenamefont {Kuhnel}(2020)}]{Carr:2020xqk}%
  \BibitemOpen
  \bibfield  {author} {\bibinfo {author} {\bibfnamefont {B.}~\bibnamefont {Carr}}\ and\ \bibinfo {author} {\bibfnamefont {F.}~\bibnamefont {Kuhnel}},\ }\href {\doibase 10.1146/annurev-nucl-050520-125911} {\bibfield  {journal} {\bibinfo  {journal} {Ann. Rev. Nucl. Part. Sci.}\ }\textbf {\bibinfo {volume} {70}},\ \bibinfo {pages} {355} (\bibinfo {year} {2020})},\ \Eprint {http://arxiv.org/abs/2006.02838} {arXiv:2006.02838 [astro-ph.CO]} \BibitemShut {NoStop}%
\bibitem [{\citenamefont {Sasaki}\ \emph {et~al.}(2016)\citenamefont {Sasaki}, \citenamefont {Suyama}, \citenamefont {Tanaka},\ and\ \citenamefont {Yokoyama}}]{Sasaki:2016jop}%
  \BibitemOpen
  \bibfield  {author} {\bibinfo {author} {\bibfnamefont {M.}~\bibnamefont {Sasaki}}, \bibinfo {author} {\bibfnamefont {T.}~\bibnamefont {Suyama}}, \bibinfo {author} {\bibfnamefont {T.}~\bibnamefont {Tanaka}}, \ and\ \bibinfo {author} {\bibfnamefont {S.}~\bibnamefont {Yokoyama}},\ }\href {\doibase 10.1103/PhysRevLett.117.061101} {\bibfield  {journal} {\bibinfo  {journal} {Phys. Rev. Lett.}\ }\textbf {\bibinfo {volume} {117}},\ \bibinfo {pages} {061101} (\bibinfo {year} {2016})},\ \bibinfo {note} {[Erratum: Phys.Rev.Lett. 121, 059901 (2018)]},\ \Eprint {http://arxiv.org/abs/1603.08338} {arXiv:1603.08338 [astro-ph.CO]} \BibitemShut {NoStop}%
\bibitem [{\citenamefont {Raidal}\ \emph {et~al.}(2017)\citenamefont {Raidal}, \citenamefont {Vaskonen},\ and\ \citenamefont {Veerm\"ae}}]{Raidal:2017mfl}%
  \BibitemOpen
  \bibfield  {author} {\bibinfo {author} {\bibfnamefont {M.}~\bibnamefont {Raidal}}, \bibinfo {author} {\bibfnamefont {V.}~\bibnamefont {Vaskonen}}, \ and\ \bibinfo {author} {\bibfnamefont {H.}~\bibnamefont {Veerm\"ae}},\ }\href {\doibase 10.1088/1475-7516/2017/09/037} {\bibfield  {journal} {\bibinfo  {journal} {JCAP}\ }\textbf {\bibinfo {volume} {09}},\ \bibinfo {pages} {037} (\bibinfo {year} {2017})},\ \Eprint {http://arxiv.org/abs/1707.01480} {arXiv:1707.01480 [astro-ph.CO]} \BibitemShut {NoStop}%
\bibitem [{\citenamefont {Ali-Ha\"\i{}moud}\ \emph {et~al.}(2017)\citenamefont {Ali-Ha\"\i{}moud}, \citenamefont {Kovetz},\ and\ \citenamefont {Kamionkowski}}]{Ali-Haimoud:2017rtz}%
  \BibitemOpen
  \bibfield  {author} {\bibinfo {author} {\bibfnamefont {Y.}~\bibnamefont {Ali-Ha\"\i{}moud}}, \bibinfo {author} {\bibfnamefont {E.~D.}\ \bibnamefont {Kovetz}}, \ and\ \bibinfo {author} {\bibfnamefont {M.}~\bibnamefont {Kamionkowski}},\ }\href {\doibase 10.1103/PhysRevD.96.123523} {\bibfield  {journal} {\bibinfo  {journal} {Phys. Rev. D}\ }\textbf {\bibinfo {volume} {96}},\ \bibinfo {pages} {123523} (\bibinfo {year} {2017})},\ \Eprint {http://arxiv.org/abs/1709.06576} {arXiv:1709.06576 [astro-ph.CO]} \BibitemShut {NoStop}%
\bibitem [{\citenamefont {Raidal}\ \emph {et~al.}(2019)\citenamefont {Raidal}, \citenamefont {Spethmann}, \citenamefont {Vaskonen},\ and\ \citenamefont {Veerm\"ae}}]{Raidal:2018bbj}%
  \BibitemOpen
  \bibfield  {author} {\bibinfo {author} {\bibfnamefont {M.}~\bibnamefont {Raidal}}, \bibinfo {author} {\bibfnamefont {C.}~\bibnamefont {Spethmann}}, \bibinfo {author} {\bibfnamefont {V.}~\bibnamefont {Vaskonen}}, \ and\ \bibinfo {author} {\bibfnamefont {H.}~\bibnamefont {Veerm\"ae}},\ }\href {\doibase 10.1088/1475-7516/2019/02/018} {\bibfield  {journal} {\bibinfo  {journal} {JCAP}\ }\textbf {\bibinfo {volume} {02}},\ \bibinfo {pages} {018} (\bibinfo {year} {2019})},\ \Eprint {http://arxiv.org/abs/1812.01930} {arXiv:1812.01930 [astro-ph.CO]} \BibitemShut {NoStop}%
\bibitem [{\citenamefont {Vaskonen}\ and\ \citenamefont {Veerm\"ae}(2020)}]{Vaskonen:2019jpv}%
  \BibitemOpen
  \bibfield  {author} {\bibinfo {author} {\bibfnamefont {V.}~\bibnamefont {Vaskonen}}\ and\ \bibinfo {author} {\bibfnamefont {H.}~\bibnamefont {Veerm\"ae}},\ }\href {\doibase 10.1103/PhysRevD.101.043015} {\bibfield  {journal} {\bibinfo  {journal} {Phys. Rev. D}\ }\textbf {\bibinfo {volume} {101}},\ \bibinfo {pages} {043015} (\bibinfo {year} {2020})},\ \Eprint {http://arxiv.org/abs/1908.09752} {arXiv:1908.09752 [astro-ph.CO]} \BibitemShut {NoStop}%
\bibitem [{\citenamefont {Hall}\ \emph {et~al.}(2020)\citenamefont {Hall}, \citenamefont {Gow},\ and\ \citenamefont {Byrnes}}]{Hall:2020daa}%
  \BibitemOpen
  \bibfield  {author} {\bibinfo {author} {\bibfnamefont {A.}~\bibnamefont {Hall}}, \bibinfo {author} {\bibfnamefont {A.~D.}\ \bibnamefont {Gow}}, \ and\ \bibinfo {author} {\bibfnamefont {C.~T.}\ \bibnamefont {Byrnes}},\ }\href {\doibase 10.1103/PhysRevD.102.123524} {\bibfield  {journal} {\bibinfo  {journal} {Phys. Rev. D}\ }\textbf {\bibinfo {volume} {102}},\ \bibinfo {pages} {123524} (\bibinfo {year} {2020})},\ \Eprint {http://arxiv.org/abs/2008.13704} {arXiv:2008.13704 [astro-ph.CO]} \BibitemShut {NoStop}%
\bibitem [{\citenamefont {Mukhanov}\ and\ \citenamefont {Chibisov}(1981)}]{Mukhanov:1981xt}%
  \BibitemOpen
  \bibfield  {author} {\bibinfo {author} {\bibfnamefont {V.~F.}\ \bibnamefont {Mukhanov}}\ and\ \bibinfo {author} {\bibfnamefont {G.~V.}\ \bibnamefont {Chibisov}},\ }\href@noop {} {\bibfield  {journal} {\bibinfo  {journal} {JETP Lett.}\ }\textbf {\bibinfo {volume} {33}},\ \bibinfo {pages} {532} (\bibinfo {year} {1981})}\BibitemShut {NoStop}%
\bibitem [{\citenamefont {Hawking}(1982)}]{Hawking:1982cz}%
  \BibitemOpen
  \bibfield  {author} {\bibinfo {author} {\bibfnamefont {S.~W.}\ \bibnamefont {Hawking}},\ }\href {\doibase 10.1016/0370-2693(82)90373-2} {\bibfield  {journal} {\bibinfo  {journal} {Phys. Lett. B}\ }\textbf {\bibinfo {volume} {115}},\ \bibinfo {pages} {295} (\bibinfo {year} {1982})}\BibitemShut {NoStop}%
\bibitem [{\citenamefont {Guth}\ and\ \citenamefont {Pi}(1982)}]{Guth:1982ec}%
  \BibitemOpen
  \bibfield  {author} {\bibinfo {author} {\bibfnamefont {A.~H.}\ \bibnamefont {Guth}}\ and\ \bibinfo {author} {\bibfnamefont {S.~Y.}\ \bibnamefont {Pi}},\ }\href {\doibase 10.1103/PhysRevLett.49.1110} {\bibfield  {journal} {\bibinfo  {journal} {Phys. Rev. Lett.}\ }\textbf {\bibinfo {volume} {49}},\ \bibinfo {pages} {1110} (\bibinfo {year} {1982})}\BibitemShut {NoStop}%
\bibitem [{\citenamefont {Starobinsky}(1982)}]{Starobinsky:1982ee}%
  \BibitemOpen
  \bibfield  {author} {\bibinfo {author} {\bibfnamefont {A.~A.}\ \bibnamefont {Starobinsky}},\ }\href {\doibase 10.1016/0370-2693(82)90541-X} {\bibfield  {journal} {\bibinfo  {journal} {Phys. Lett. B}\ }\textbf {\bibinfo {volume} {117}},\ \bibinfo {pages} {175} (\bibinfo {year} {1982})}\BibitemShut {NoStop}%
\bibitem [{\citenamefont {Starobinsky}(1980)}]{Starobinsky:1980te}%
  \BibitemOpen
  \bibfield  {author} {\bibinfo {author} {\bibfnamefont {A.~A.}\ \bibnamefont {Starobinsky}},\ }\href {\doibase 10.1016/0370-2693(80)90670-X} {\bibfield  {journal} {\bibinfo  {journal} {Phys. Lett. B}\ }\textbf {\bibinfo {volume} {91}},\ \bibinfo {pages} {99} (\bibinfo {year} {1980})}\BibitemShut {NoStop}%
\bibitem [{\citenamefont {Sato}(1981)}]{Sato:1980yn}%
  \BibitemOpen
  \bibfield  {author} {\bibinfo {author} {\bibfnamefont {K.}~\bibnamefont {Sato}},\ }\href@noop {} {\bibfield  {journal} {\bibinfo  {journal} {Mon. Not. Roy. Astron. Soc.}\ }\textbf {\bibinfo {volume} {195}},\ \bibinfo {pages} {467} (\bibinfo {year} {1981})}\BibitemShut {NoStop}%
\bibitem [{\citenamefont {Guth}(1981)}]{Guth:1980zm}%
  \BibitemOpen
  \bibfield  {author} {\bibinfo {author} {\bibfnamefont {A.~H.}\ \bibnamefont {Guth}},\ }\href {\doibase 10.1103/PhysRevD.23.347} {\bibfield  {journal} {\bibinfo  {journal} {Phys. Rev. D}\ }\textbf {\bibinfo {volume} {23}},\ \bibinfo {pages} {347} (\bibinfo {year} {1981})}\BibitemShut {NoStop}%
\bibitem [{\citenamefont {Aghanim}\ \emph {et~al.}(2020)\citenamefont {Aghanim} \emph {et~al.}}]{Planck:2018nkj}%
  \BibitemOpen
  \bibfield  {author} {\bibinfo {author} {\bibfnamefont {N.}~\bibnamefont {Aghanim}} \emph {et~al.} (\bibinfo {collaboration} {Planck}),\ }\href {\doibase 10.1051/0004-6361/201833880} {\bibfield  {journal} {\bibinfo  {journal} {Astron. Astrophys.}\ }\textbf {\bibinfo {volume} {641}},\ \bibinfo {pages} {A1} (\bibinfo {year} {2020})},\ \Eprint {http://arxiv.org/abs/1807.06205} {arXiv:1807.06205 [astro-ph.CO]} \BibitemShut {NoStop}%
\bibitem [{\citenamefont {Akrami}\ \emph {et~al.}(2020{\natexlab{a}})\citenamefont {Akrami} \emph {et~al.}}]{Planck:2018jri}%
  \BibitemOpen
  \bibfield  {author} {\bibinfo {author} {\bibfnamefont {Y.}~\bibnamefont {Akrami}} \emph {et~al.} (\bibinfo {collaboration} {Planck}),\ }\href {\doibase 10.1051/0004-6361/201833887} {\bibfield  {journal} {\bibinfo  {journal} {Astron. Astrophys.}\ }\textbf {\bibinfo {volume} {641}},\ \bibinfo {pages} {A10} (\bibinfo {year} {2020}{\natexlab{a}})},\ \Eprint {http://arxiv.org/abs/1807.06211} {arXiv:1807.06211 [astro-ph.CO]} \BibitemShut {NoStop}%
\bibitem [{\citenamefont {Akrami}\ \emph {et~al.}(2020{\natexlab{b}})\citenamefont {Akrami} \emph {et~al.}}]{Planck:2019kim}%
  \BibitemOpen
  \bibfield  {author} {\bibinfo {author} {\bibfnamefont {Y.}~\bibnamefont {Akrami}} \emph {et~al.} (\bibinfo {collaboration} {Planck}),\ }\href {\doibase 10.1051/0004-6361/201935891} {\bibfield  {journal} {\bibinfo  {journal} {Astron. Astrophys.}\ }\textbf {\bibinfo {volume} {641}},\ \bibinfo {pages} {A9} (\bibinfo {year} {2020}{\natexlab{b}})},\ \Eprint {http://arxiv.org/abs/1905.05697} {arXiv:1905.05697 [astro-ph.CO]} \BibitemShut {NoStop}%
\bibitem [{\citenamefont {Nakama}\ \emph {et~al.}(2014)\citenamefont {Nakama}, \citenamefont {Suyama},\ and\ \citenamefont {Yokoyama}}]{Nakama:2014vla}%
  \BibitemOpen
  \bibfield  {author} {\bibinfo {author} {\bibfnamefont {T.}~\bibnamefont {Nakama}}, \bibinfo {author} {\bibfnamefont {T.}~\bibnamefont {Suyama}}, \ and\ \bibinfo {author} {\bibfnamefont {J.}~\bibnamefont {Yokoyama}},\ }\href {\doibase 10.1103/PhysRevLett.113.061302} {\bibfield  {journal} {\bibinfo  {journal} {Phys. Rev. Lett.}\ }\textbf {\bibinfo {volume} {113}},\ \bibinfo {pages} {061302} (\bibinfo {year} {2014})},\ \Eprint {http://arxiv.org/abs/1403.5407} {arXiv:1403.5407 [astro-ph.CO]} \BibitemShut {NoStop}%
\bibitem [{\citenamefont {Jeong}\ \emph {et~al.}(2014)\citenamefont {Jeong}, \citenamefont {Pradler}, \citenamefont {Chluba},\ and\ \citenamefont {Kamionkowski}}]{Jeong:2014gna}%
  \BibitemOpen
  \bibfield  {author} {\bibinfo {author} {\bibfnamefont {D.}~\bibnamefont {Jeong}}, \bibinfo {author} {\bibfnamefont {J.}~\bibnamefont {Pradler}}, \bibinfo {author} {\bibfnamefont {J.}~\bibnamefont {Chluba}}, \ and\ \bibinfo {author} {\bibfnamefont {M.}~\bibnamefont {Kamionkowski}},\ }\href {\doibase 10.1103/PhysRevLett.113.061301} {\bibfield  {journal} {\bibinfo  {journal} {Phys. Rev. Lett.}\ }\textbf {\bibinfo {volume} {113}},\ \bibinfo {pages} {061301} (\bibinfo {year} {2014})},\ \Eprint {http://arxiv.org/abs/1403.3697} {arXiv:1403.3697 [astro-ph.CO]} \BibitemShut {NoStop}%
\bibitem [{\citenamefont {Inomata}\ \emph {et~al.}(2016)\citenamefont {Inomata}, \citenamefont {Kawasaki},\ and\ \citenamefont {Tada}}]{Inomata:2016uip}%
  \BibitemOpen
  \bibfield  {author} {\bibinfo {author} {\bibfnamefont {K.}~\bibnamefont {Inomata}}, \bibinfo {author} {\bibfnamefont {M.}~\bibnamefont {Kawasaki}}, \ and\ \bibinfo {author} {\bibfnamefont {Y.}~\bibnamefont {Tada}},\ }\href {\doibase 10.1103/PhysRevD.94.043527} {\bibfield  {journal} {\bibinfo  {journal} {Phys. Rev. D}\ }\textbf {\bibinfo {volume} {94}},\ \bibinfo {pages} {043527} (\bibinfo {year} {2016})},\ \Eprint {http://arxiv.org/abs/1605.04646} {arXiv:1605.04646 [astro-ph.CO]} \BibitemShut {NoStop}%
\bibitem [{\citenamefont {Nakama}\ \emph {et~al.}(2018)\citenamefont {Nakama}, \citenamefont {Suyama}, \citenamefont {Kohri},\ and\ \citenamefont {Hiroshima}}]{Nakama:2017qac}%
  \BibitemOpen
  \bibfield  {author} {\bibinfo {author} {\bibfnamefont {T.}~\bibnamefont {Nakama}}, \bibinfo {author} {\bibfnamefont {T.}~\bibnamefont {Suyama}}, \bibinfo {author} {\bibfnamefont {K.}~\bibnamefont {Kohri}}, \ and\ \bibinfo {author} {\bibfnamefont {N.}~\bibnamefont {Hiroshima}},\ }\href {\doibase 10.1103/PhysRevD.97.023539} {\bibfield  {journal} {\bibinfo  {journal} {Phys. Rev. D}\ }\textbf {\bibinfo {volume} {97}},\ \bibinfo {pages} {023539} (\bibinfo {year} {2018})},\ \Eprint {http://arxiv.org/abs/1712.08820} {arXiv:1712.08820 [astro-ph.CO]} \BibitemShut {NoStop}%
\bibitem [{\citenamefont {Kawasaki}\ \emph {et~al.}(2022)\citenamefont {Kawasaki}, \citenamefont {Nakatsuka},\ and\ \citenamefont {Nakayama}}]{Kawasaki:2021yek}%
  \BibitemOpen
  \bibfield  {author} {\bibinfo {author} {\bibfnamefont {M.}~\bibnamefont {Kawasaki}}, \bibinfo {author} {\bibfnamefont {H.}~\bibnamefont {Nakatsuka}}, \ and\ \bibinfo {author} {\bibfnamefont {K.}~\bibnamefont {Nakayama}},\ }\href {\doibase 10.1088/1475-7516/2022/03/061} {\bibfield  {journal} {\bibinfo  {journal} {JCAP}\ }\textbf {\bibinfo {volume} {03}},\ \bibinfo {pages} {061} (\bibinfo {year} {2022})},\ \Eprint {http://arxiv.org/abs/2110.12620} {arXiv:2110.12620 [astro-ph.CO]} \BibitemShut {NoStop}%
\bibitem [{\citenamefont {Kimura}\ \emph {et~al.}(2021)\citenamefont {Kimura}, \citenamefont {Suyama}, \citenamefont {Yamaguchi},\ and\ \citenamefont {Zhang}}]{Kimura:2021sqz}%
  \BibitemOpen
  \bibfield  {author} {\bibinfo {author} {\bibfnamefont {R.}~\bibnamefont {Kimura}}, \bibinfo {author} {\bibfnamefont {T.}~\bibnamefont {Suyama}}, \bibinfo {author} {\bibfnamefont {M.}~\bibnamefont {Yamaguchi}}, \ and\ \bibinfo {author} {\bibfnamefont {Y.-L.}\ \bibnamefont {Zhang}},\ }\href {\doibase 10.1088/1475-7516/2021/04/031} {\bibfield  {journal} {\bibinfo  {journal} {JCAP}\ }\textbf {\bibinfo {volume} {04}},\ \bibinfo {pages} {031} (\bibinfo {year} {2021})},\ \Eprint {http://arxiv.org/abs/2102.05280} {arXiv:2102.05280 [astro-ph.CO]} \BibitemShut {NoStop}%
\bibitem [{\citenamefont {Wang}\ \emph {et~al.}(2023)\citenamefont {Wang}, \citenamefont {Zhang}, \citenamefont {Kimura},\ and\ \citenamefont {Yamaguchi}}]{Wang:2022nml}%
  \BibitemOpen
  \bibfield  {author} {\bibinfo {author} {\bibfnamefont {X.}~\bibnamefont {Wang}}, \bibinfo {author} {\bibfnamefont {Y.-l.}\ \bibnamefont {Zhang}}, \bibinfo {author} {\bibfnamefont {R.}~\bibnamefont {Kimura}}, \ and\ \bibinfo {author} {\bibfnamefont {M.}~\bibnamefont {Yamaguchi}},\ }\href {\doibase 10.1007/s11433-023-2091-x} {\bibfield  {journal} {\bibinfo  {journal} {Sci. China Phys. Mech. Astron.}\ }\textbf {\bibinfo {volume} {66}},\ \bibinfo {pages} {260462} (\bibinfo {year} {2023})},\ \Eprint {http://arxiv.org/abs/2209.12911} {arXiv:2209.12911 [astro-ph.CO]} \BibitemShut {NoStop}%
\bibitem [{\citenamefont {Carr}\ and\ \citenamefont {Lidsey}(1993)}]{Carr:1993aq}%
  \BibitemOpen
  \bibfield  {author} {\bibinfo {author} {\bibfnamefont {B.~J.}\ \bibnamefont {Carr}}\ and\ \bibinfo {author} {\bibfnamefont {J.~E.}\ \bibnamefont {Lidsey}},\ }\href {\doibase 10.1103/PhysRevD.48.543} {\bibfield  {journal} {\bibinfo  {journal} {Phys. Rev. D}\ }\textbf {\bibinfo {volume} {48}},\ \bibinfo {pages} {543} (\bibinfo {year} {1993})}\BibitemShut {NoStop}%
\bibitem [{\citenamefont {Yokoyama}(1998)}]{Yokoyama:1998pt}%
  \BibitemOpen
  \bibfield  {author} {\bibinfo {author} {\bibfnamefont {J.}~\bibnamefont {Yokoyama}},\ }\href {\doibase 10.1103/PhysRevD.58.083510} {\bibfield  {journal} {\bibinfo  {journal} {Phys. Rev. D}\ }\textbf {\bibinfo {volume} {58}},\ \bibinfo {pages} {083510} (\bibinfo {year} {1998})},\ \Eprint {http://arxiv.org/abs/astro-ph/9802357} {arXiv:astro-ph/9802357} \BibitemShut {NoStop}%
\bibitem [{\citenamefont {Kawasaki}\ \emph {et~al.}(1998)\citenamefont {Kawasaki}, \citenamefont {Sugiyama},\ and\ \citenamefont {Yanagida}}]{Kawasaki:1997ju}%
  \BibitemOpen
  \bibfield  {author} {\bibinfo {author} {\bibfnamefont {M.}~\bibnamefont {Kawasaki}}, \bibinfo {author} {\bibfnamefont {N.}~\bibnamefont {Sugiyama}}, \ and\ \bibinfo {author} {\bibfnamefont {T.}~\bibnamefont {Yanagida}},\ }\href {\doibase 10.1103/PhysRevD.57.6050} {\bibfield  {journal} {\bibinfo  {journal} {Phys. Rev. D}\ }\textbf {\bibinfo {volume} {57}},\ \bibinfo {pages} {6050} (\bibinfo {year} {1998})},\ \Eprint {http://arxiv.org/abs/hep-ph/9710259} {arXiv:hep-ph/9710259} \BibitemShut {NoStop}%
\bibitem [{\citenamefont {Kawasaki}\ and\ \citenamefont {Yanagida}(1999)}]{Kawasaki:1998vx}%
  \BibitemOpen
  \bibfield  {author} {\bibinfo {author} {\bibfnamefont {M.}~\bibnamefont {Kawasaki}}\ and\ \bibinfo {author} {\bibfnamefont {T.}~\bibnamefont {Yanagida}},\ }\href {\doibase 10.1103/PhysRevD.59.043512} {\bibfield  {journal} {\bibinfo  {journal} {Phys. Rev. D}\ }\textbf {\bibinfo {volume} {59}},\ \bibinfo {pages} {043512} (\bibinfo {year} {1999})},\ \Eprint {http://arxiv.org/abs/hep-ph/9807544} {arXiv:hep-ph/9807544} \BibitemShut {NoStop}%
\bibitem [{\citenamefont {Kawasaki}\ \emph {et~al.}(2016)\citenamefont {Kawasaki}, \citenamefont {Kusenko}, \citenamefont {Tada},\ and\ \citenamefont {Yanagida}}]{Kawasaki:2016pql}%
  \BibitemOpen
  \bibfield  {author} {\bibinfo {author} {\bibfnamefont {M.}~\bibnamefont {Kawasaki}}, \bibinfo {author} {\bibfnamefont {A.}~\bibnamefont {Kusenko}}, \bibinfo {author} {\bibfnamefont {Y.}~\bibnamefont {Tada}}, \ and\ \bibinfo {author} {\bibfnamefont {T.~T.}\ \bibnamefont {Yanagida}},\ }\href {\doibase 10.1103/PhysRevD.94.083523} {\bibfield  {journal} {\bibinfo  {journal} {Phys. Rev. D}\ }\textbf {\bibinfo {volume} {94}},\ \bibinfo {pages} {083523} (\bibinfo {year} {2016})},\ \Eprint {http://arxiv.org/abs/1606.07631} {arXiv:1606.07631 [astro-ph.CO]} \BibitemShut {NoStop}%
\bibitem [{\citenamefont {Gao}\ and\ \citenamefont {Guo}(2018)}]{Gao:2018pvq}%
  \BibitemOpen
  \bibfield  {author} {\bibinfo {author} {\bibfnamefont {T.-J.}\ \bibnamefont {Gao}}\ and\ \bibinfo {author} {\bibfnamefont {Z.-K.}\ \bibnamefont {Guo}},\ }\href {\doibase 10.1103/PhysRevD.98.063526} {\bibfield  {journal} {\bibinfo  {journal} {Phys. Rev. D}\ }\textbf {\bibinfo {volume} {98}},\ \bibinfo {pages} {063526} (\bibinfo {year} {2018})},\ \Eprint {http://arxiv.org/abs/1806.09320} {arXiv:1806.09320 [hep-ph]} \BibitemShut {NoStop}%
\bibitem [{\citenamefont {Nanopoulos}\ \emph {et~al.}(2020)\citenamefont {Nanopoulos}, \citenamefont {Spanos},\ and\ \citenamefont {Stamou}}]{Nanopoulos:2020nnh}%
  \BibitemOpen
  \bibfield  {author} {\bibinfo {author} {\bibfnamefont {D.~V.}\ \bibnamefont {Nanopoulos}}, \bibinfo {author} {\bibfnamefont {V.~C.}\ \bibnamefont {Spanos}}, \ and\ \bibinfo {author} {\bibfnamefont {I.~D.}\ \bibnamefont {Stamou}},\ }\href {\doibase 10.1103/PhysRevD.102.083536} {\bibfield  {journal} {\bibinfo  {journal} {Phys. Rev. D}\ }\textbf {\bibinfo {volume} {102}},\ \bibinfo {pages} {083536} (\bibinfo {year} {2020})},\ \Eprint {http://arxiv.org/abs/2008.01457} {arXiv:2008.01457 [astro-ph.CO]} \BibitemShut {NoStop}%
\bibitem [{\citenamefont {Wu}\ \emph {et~al.}(2021)\citenamefont {Wu}, \citenamefont {Gong},\ and\ \citenamefont {Li}}]{Wu:2021zta}%
  \BibitemOpen
  \bibfield  {author} {\bibinfo {author} {\bibfnamefont {L.}~\bibnamefont {Wu}}, \bibinfo {author} {\bibfnamefont {Y.}~\bibnamefont {Gong}}, \ and\ \bibinfo {author} {\bibfnamefont {T.}~\bibnamefont {Li}},\ }\href {\doibase 10.1103/PhysRevD.104.123544} {\bibfield  {journal} {\bibinfo  {journal} {Phys. Rev. D}\ }\textbf {\bibinfo {volume} {104}},\ \bibinfo {pages} {123544} (\bibinfo {year} {2021})},\ \Eprint {http://arxiv.org/abs/2105.07694} {arXiv:2105.07694 [gr-qc]} \BibitemShut {NoStop}%
\bibitem [{\citenamefont {Stamou}(2021)}]{Stamou:2021qdk}%
  \BibitemOpen
  \bibfield  {author} {\bibinfo {author} {\bibfnamefont {I.~D.}\ \bibnamefont {Stamou}},\ }\href {\doibase 10.1103/PhysRevD.103.083512} {\bibfield  {journal} {\bibinfo  {journal} {Phys. Rev. D}\ }\textbf {\bibinfo {volume} {103}},\ \bibinfo {pages} {083512} (\bibinfo {year} {2021})},\ \Eprint {http://arxiv.org/abs/2104.08654} {arXiv:2104.08654 [hep-ph]} \BibitemShut {NoStop}%
\bibitem [{\citenamefont {Cheng}\ \emph {et~al.}(2018)\citenamefont {Cheng}, \citenamefont {Lee},\ and\ \citenamefont {Ng}}]{Cheng:2018yyr}%
  \BibitemOpen
  \bibfield  {author} {\bibinfo {author} {\bibfnamefont {S.-L.}\ \bibnamefont {Cheng}}, \bibinfo {author} {\bibfnamefont {W.}~\bibnamefont {Lee}}, \ and\ \bibinfo {author} {\bibfnamefont {K.-W.}\ \bibnamefont {Ng}},\ }\href {\doibase 10.1088/1475-7516/2018/07/001} {\bibfield  {journal} {\bibinfo  {journal} {JCAP}\ }\textbf {\bibinfo {volume} {07}},\ \bibinfo {pages} {001} (\bibinfo {year} {2018})},\ \Eprint {http://arxiv.org/abs/1801.09050} {arXiv:1801.09050 [astro-ph.CO]} \BibitemShut {NoStop}%
\bibitem [{\citenamefont {Ballesteros}\ \emph {et~al.}(2020{\natexlab{a}})\citenamefont {Ballesteros}, \citenamefont {Rey},\ and\ \citenamefont {Rompineve}}]{Ballesteros:2019hus}%
  \BibitemOpen
  \bibfield  {author} {\bibinfo {author} {\bibfnamefont {G.}~\bibnamefont {Ballesteros}}, \bibinfo {author} {\bibfnamefont {J.}~\bibnamefont {Rey}}, \ and\ \bibinfo {author} {\bibfnamefont {F.}~\bibnamefont {Rompineve}},\ }\href {\doibase 10.1088/1475-7516/2020/06/014} {\bibfield  {journal} {\bibinfo  {journal} {JCAP}\ }\textbf {\bibinfo {volume} {06}},\ \bibinfo {pages} {014} (\bibinfo {year} {2020}{\natexlab{a}})},\ \Eprint {http://arxiv.org/abs/1912.01638} {arXiv:1912.01638 [astro-ph.CO]} \BibitemShut {NoStop}%
\bibitem [{\citenamefont {Pi}\ \emph {et~al.}(2018)\citenamefont {Pi}, \citenamefont {Zhang}, \citenamefont {Huang},\ and\ \citenamefont {Sasaki}}]{Pi:2017gih}%
  \BibitemOpen
  \bibfield  {author} {\bibinfo {author} {\bibfnamefont {S.}~\bibnamefont {Pi}}, \bibinfo {author} {\bibfnamefont {Y.-l.}\ \bibnamefont {Zhang}}, \bibinfo {author} {\bibfnamefont {Q.-G.}\ \bibnamefont {Huang}}, \ and\ \bibinfo {author} {\bibfnamefont {M.}~\bibnamefont {Sasaki}},\ }\href {\doibase 10.1088/1475-7516/2018/05/042} {\bibfield  {journal} {\bibinfo  {journal} {JCAP}\ }\textbf {\bibinfo {volume} {05}},\ \bibinfo {pages} {042} (\bibinfo {year} {2018})},\ \Eprint {http://arxiv.org/abs/1712.09896} {arXiv:1712.09896 [astro-ph.CO]} \BibitemShut {NoStop}%
\bibitem [{\citenamefont {Dalianis}\ and\ \citenamefont {Tringas}(2019)}]{Dalianis:2019asr}%
  \BibitemOpen
  \bibfield  {author} {\bibinfo {author} {\bibfnamefont {I.}~\bibnamefont {Dalianis}}\ and\ \bibinfo {author} {\bibfnamefont {G.}~\bibnamefont {Tringas}},\ }\href {\doibase 10.1103/PhysRevD.100.083512} {\bibfield  {journal} {\bibinfo  {journal} {Phys. Rev. D}\ }\textbf {\bibinfo {volume} {100}},\ \bibinfo {pages} {083512} (\bibinfo {year} {2019})},\ \Eprint {http://arxiv.org/abs/1905.01741} {arXiv:1905.01741 [astro-ph.CO]} \BibitemShut {NoStop}%
\bibitem [{\citenamefont {Dalianis}\ \emph {et~al.}(2019)\citenamefont {Dalianis}, \citenamefont {Kehagias},\ and\ \citenamefont {Tringas}}]{Dalianis:2018frf}%
  \BibitemOpen
  \bibfield  {author} {\bibinfo {author} {\bibfnamefont {I.}~\bibnamefont {Dalianis}}, \bibinfo {author} {\bibfnamefont {A.}~\bibnamefont {Kehagias}}, \ and\ \bibinfo {author} {\bibfnamefont {G.}~\bibnamefont {Tringas}},\ }\href {\doibase 10.1088/1475-7516/2019/01/037} {\bibfield  {journal} {\bibinfo  {journal} {JCAP}\ }\textbf {\bibinfo {volume} {01}},\ \bibinfo {pages} {037} (\bibinfo {year} {2019})},\ \Eprint {http://arxiv.org/abs/1805.09483} {arXiv:1805.09483 [astro-ph.CO]} \BibitemShut {NoStop}%
\bibitem [{\citenamefont {Mahbub}(2020)}]{Mahbub:2019uhl}%
  \BibitemOpen
  \bibfield  {author} {\bibinfo {author} {\bibfnamefont {R.}~\bibnamefont {Mahbub}},\ }\href {\doibase 10.1103/PhysRevD.101.023533} {\bibfield  {journal} {\bibinfo  {journal} {Phys. Rev. D}\ }\textbf {\bibinfo {volume} {101}},\ \bibinfo {pages} {023533} (\bibinfo {year} {2020})},\ \Eprint {http://arxiv.org/abs/1910.10602} {arXiv:1910.10602 [astro-ph.CO]} \BibitemShut {NoStop}%
\bibitem [{\citenamefont {Cicoli}\ \emph {et~al.}(2018)\citenamefont {Cicoli}, \citenamefont {Diaz},\ and\ \citenamefont {Pedro}}]{Cicoli:2018asa}%
  \BibitemOpen
  \bibfield  {author} {\bibinfo {author} {\bibfnamefont {M.}~\bibnamefont {Cicoli}}, \bibinfo {author} {\bibfnamefont {V.~A.}\ \bibnamefont {Diaz}}, \ and\ \bibinfo {author} {\bibfnamefont {F.~G.}\ \bibnamefont {Pedro}},\ }\href {\doibase 10.1088/1475-7516/2018/06/034} {\bibfield  {journal} {\bibinfo  {journal} {JCAP}\ }\textbf {\bibinfo {volume} {06}},\ \bibinfo {pages} {034} (\bibinfo {year} {2018})},\ \Eprint {http://arxiv.org/abs/1803.02837} {arXiv:1803.02837 [hep-th]} \BibitemShut {NoStop}%
\bibitem [{\citenamefont {\"Ozsoy}\ \emph {et~al.}(2018)\citenamefont {\"Ozsoy}, \citenamefont {Parameswaran}, \citenamefont {Tasinato},\ and\ \citenamefont {Zavala}}]{Ozsoy:2018flq}%
  \BibitemOpen
  \bibfield  {author} {\bibinfo {author} {\bibfnamefont {O.}~\bibnamefont {\"Ozsoy}}, \bibinfo {author} {\bibfnamefont {S.}~\bibnamefont {Parameswaran}}, \bibinfo {author} {\bibfnamefont {G.}~\bibnamefont {Tasinato}}, \ and\ \bibinfo {author} {\bibfnamefont {I.}~\bibnamefont {Zavala}},\ }\href {\doibase 10.1088/1475-7516/2018/07/005} {\bibfield  {journal} {\bibinfo  {journal} {JCAP}\ }\textbf {\bibinfo {volume} {07}},\ \bibinfo {pages} {005} (\bibinfo {year} {2018})},\ \Eprint {http://arxiv.org/abs/1803.07626} {arXiv:1803.07626 [hep-th]} \BibitemShut {NoStop}%
\bibitem [{\citenamefont {Cicoli}\ \emph {et~al.}(2022)\citenamefont {Cicoli}, \citenamefont {Pedro},\ and\ \citenamefont {Pedron}}]{Cicoli:2022sih}%
  \BibitemOpen
  \bibfield  {author} {\bibinfo {author} {\bibfnamefont {M.}~\bibnamefont {Cicoli}}, \bibinfo {author} {\bibfnamefont {F.~G.}\ \bibnamefont {Pedro}}, \ and\ \bibinfo {author} {\bibfnamefont {N.}~\bibnamefont {Pedron}},\ }\href {\doibase 10.1088/1475-7516/2022/08/030} {\bibfield  {journal} {\bibinfo  {journal} {JCAP}\ }\textbf {\bibinfo {volume} {08}},\ \bibinfo {pages} {030} (\bibinfo {year} {2022})},\ \Eprint {http://arxiv.org/abs/2203.00021} {arXiv:2203.00021 [hep-th]} \BibitemShut {NoStop}%
\bibitem [{\citenamefont {Ezquiaga}\ \emph {et~al.}(2018)\citenamefont {Ezquiaga}, \citenamefont {Garcia-Bellido},\ and\ \citenamefont {Ruiz~Morales}}]{Ezquiaga:2017fvi}%
  \BibitemOpen
  \bibfield  {author} {\bibinfo {author} {\bibfnamefont {J.~M.}\ \bibnamefont {Ezquiaga}}, \bibinfo {author} {\bibfnamefont {J.}~\bibnamefont {Garcia-Bellido}}, \ and\ \bibinfo {author} {\bibfnamefont {E.}~\bibnamefont {Ruiz~Morales}},\ }\href {\doibase 10.1016/j.physletb.2017.11.039} {\bibfield  {journal} {\bibinfo  {journal} {Phys. Lett. B}\ }\textbf {\bibinfo {volume} {776}},\ \bibinfo {pages} {345} (\bibinfo {year} {2018})},\ \Eprint {http://arxiv.org/abs/1705.04861} {arXiv:1705.04861 [astro-ph.CO]} \BibitemShut {NoStop}%
\bibitem [{\citenamefont {Ballesteros}\ and\ \citenamefont {Taoso}(2018)}]{Ballesteros:2017fsr}%
  \BibitemOpen
  \bibfield  {author} {\bibinfo {author} {\bibfnamefont {G.}~\bibnamefont {Ballesteros}}\ and\ \bibinfo {author} {\bibfnamefont {M.}~\bibnamefont {Taoso}},\ }\href {\doibase 10.1103/PhysRevD.97.023501} {\bibfield  {journal} {\bibinfo  {journal} {Phys. Rev. D}\ }\textbf {\bibinfo {volume} {97}},\ \bibinfo {pages} {023501} (\bibinfo {year} {2018})},\ \Eprint {http://arxiv.org/abs/1709.05565} {arXiv:1709.05565 [hep-ph]} \BibitemShut {NoStop}%
\bibitem [{\citenamefont {Drees}\ and\ \citenamefont {Xu}(2021)}]{Drees:2019xpp}%
  \BibitemOpen
  \bibfield  {author} {\bibinfo {author} {\bibfnamefont {M.}~\bibnamefont {Drees}}\ and\ \bibinfo {author} {\bibfnamefont {Y.}~\bibnamefont {Xu}},\ }\href {\doibase 10.1140/epjc/s10052-021-08976-2} {\bibfield  {journal} {\bibinfo  {journal} {Eur. Phys. J. C}\ }\textbf {\bibinfo {volume} {81}},\ \bibinfo {pages} {182} (\bibinfo {year} {2021})},\ \Eprint {http://arxiv.org/abs/1905.13581} {arXiv:1905.13581 [hep-ph]} \BibitemShut {NoStop}%
\bibitem [{\citenamefont {Cheong}\ \emph {et~al.}(2021)\citenamefont {Cheong}, \citenamefont {Lee},\ and\ \citenamefont {Park}}]{Cheong:2019vzl}%
  \BibitemOpen
  \bibfield  {author} {\bibinfo {author} {\bibfnamefont {D.~Y.}\ \bibnamefont {Cheong}}, \bibinfo {author} {\bibfnamefont {S.~M.}\ \bibnamefont {Lee}}, \ and\ \bibinfo {author} {\bibfnamefont {S.~C.}\ \bibnamefont {Park}},\ }\href {\doibase 10.1088/1475-7516/2021/01/032} {\bibfield  {journal} {\bibinfo  {journal} {JCAP}\ }\textbf {\bibinfo {volume} {01}},\ \bibinfo {pages} {032} (\bibinfo {year} {2021})},\ \Eprint {http://arxiv.org/abs/1912.12032} {arXiv:1912.12032 [hep-ph]} \BibitemShut {NoStop}%
\bibitem [{\citenamefont {Rasanen}\ and\ \citenamefont {Tomberg}(2019)}]{Rasanen:2018fom}%
  \BibitemOpen
  \bibfield  {author} {\bibinfo {author} {\bibfnamefont {S.}~\bibnamefont {Rasanen}}\ and\ \bibinfo {author} {\bibfnamefont {E.}~\bibnamefont {Tomberg}},\ }\href {\doibase 10.1088/1475-7516/2019/01/038} {\bibfield  {journal} {\bibinfo  {journal} {JCAP}\ }\textbf {\bibinfo {volume} {01}},\ \bibinfo {pages} {038} (\bibinfo {year} {2019})},\ \Eprint {http://arxiv.org/abs/1810.12608} {arXiv:1810.12608 [astro-ph.CO]} \BibitemShut {NoStop}%
\bibitem [{\citenamefont {Garcia-Bellido}\ and\ \citenamefont {Ruiz~Morales}(2017)}]{Garcia-Bellido:2017mdw}%
  \BibitemOpen
  \bibfield  {author} {\bibinfo {author} {\bibfnamefont {J.}~\bibnamefont {Garcia-Bellido}}\ and\ \bibinfo {author} {\bibfnamefont {E.}~\bibnamefont {Ruiz~Morales}},\ }\href {\doibase 10.1016/j.dark.2017.09.007} {\bibfield  {journal} {\bibinfo  {journal} {Phys. Dark Univ.}\ }\textbf {\bibinfo {volume} {18}},\ \bibinfo {pages} {47} (\bibinfo {year} {2017})},\ \Eprint {http://arxiv.org/abs/1702.03901} {arXiv:1702.03901 [astro-ph.CO]} \BibitemShut {NoStop}%
\bibitem [{\citenamefont {Ragavendra}\ \emph {et~al.}(2021)\citenamefont {Ragavendra}, \citenamefont {Saha}, \citenamefont {Sriramkumar},\ and\ \citenamefont {Silk}}]{Ragavendra:2020sop}%
  \BibitemOpen
  \bibfield  {author} {\bibinfo {author} {\bibfnamefont {H.~V.}\ \bibnamefont {Ragavendra}}, \bibinfo {author} {\bibfnamefont {P.}~\bibnamefont {Saha}}, \bibinfo {author} {\bibfnamefont {L.}~\bibnamefont {Sriramkumar}}, \ and\ \bibinfo {author} {\bibfnamefont {J.}~\bibnamefont {Silk}},\ }\href {\doibase 10.1103/PhysRevD.103.083510} {\bibfield  {journal} {\bibinfo  {journal} {Phys. Rev. D}\ }\textbf {\bibinfo {volume} {103}},\ \bibinfo {pages} {083510} (\bibinfo {year} {2021})},\ \Eprint {http://arxiv.org/abs/2008.12202} {arXiv:2008.12202 [astro-ph.CO]} \BibitemShut {NoStop}%
\bibitem [{\citenamefont {Mishra}\ and\ \citenamefont {Sahni}(2020)}]{Mishra:2019pzq}%
  \BibitemOpen
  \bibfield  {author} {\bibinfo {author} {\bibfnamefont {S.~S.}\ \bibnamefont {Mishra}}\ and\ \bibinfo {author} {\bibfnamefont {V.}~\bibnamefont {Sahni}},\ }\href {\doibase 10.1088/1475-7516/2020/04/007} {\bibfield  {journal} {\bibinfo  {journal} {JCAP}\ }\textbf {\bibinfo {volume} {04}},\ \bibinfo {pages} {007} (\bibinfo {year} {2020})},\ \Eprint {http://arxiv.org/abs/1911.00057} {arXiv:1911.00057 [gr-qc]} \BibitemShut {NoStop}%
\bibitem [{\citenamefont {Atal}\ \emph {et~al.}(2019)\citenamefont {Atal}, \citenamefont {Garriga},\ and\ \citenamefont {Marcos-Caballero}}]{Atal:2019cdz}%
  \BibitemOpen
  \bibfield  {author} {\bibinfo {author} {\bibfnamefont {V.}~\bibnamefont {Atal}}, \bibinfo {author} {\bibfnamefont {J.}~\bibnamefont {Garriga}}, \ and\ \bibinfo {author} {\bibfnamefont {A.}~\bibnamefont {Marcos-Caballero}},\ }\href {\doibase 10.1088/1475-7516/2019/09/073} {\bibfield  {journal} {\bibinfo  {journal} {JCAP}\ }\textbf {\bibinfo {volume} {09}},\ \bibinfo {pages} {073} (\bibinfo {year} {2019})},\ \Eprint {http://arxiv.org/abs/1905.13202} {arXiv:1905.13202 [astro-ph.CO]} \BibitemShut {NoStop}%
\bibitem [{\citenamefont {Zheng}\ \emph {et~al.}(2022)\citenamefont {Zheng}, \citenamefont {Shi},\ and\ \citenamefont {Qiu}}]{Zheng:2021vda}%
  \BibitemOpen
  \bibfield  {author} {\bibinfo {author} {\bibfnamefont {R.}~\bibnamefont {Zheng}}, \bibinfo {author} {\bibfnamefont {J.}~\bibnamefont {Shi}}, \ and\ \bibinfo {author} {\bibfnamefont {T.}~\bibnamefont {Qiu}},\ }\href {\doibase 10.1088/1674-1137/ac42bd} {\bibfield  {journal} {\bibinfo  {journal} {Chin. Phys. C}\ }\textbf {\bibinfo {volume} {46}},\ \bibinfo {pages} {045103} (\bibinfo {year} {2022})},\ \Eprint {http://arxiv.org/abs/2106.04303} {arXiv:2106.04303 [astro-ph.CO]} \BibitemShut {NoStop}%
\bibitem [{\citenamefont {Wang}\ \emph {et~al.}(2021)\citenamefont {Wang}, \citenamefont {Liu}, \citenamefont {Su},\ and\ \citenamefont {Li}}]{Wang:2021kbh}%
  \BibitemOpen
  \bibfield  {author} {\bibinfo {author} {\bibfnamefont {Q.}~\bibnamefont {Wang}}, \bibinfo {author} {\bibfnamefont {Y.-C.}\ \bibnamefont {Liu}}, \bibinfo {author} {\bibfnamefont {B.-Y.}\ \bibnamefont {Su}}, \ and\ \bibinfo {author} {\bibfnamefont {N.}~\bibnamefont {Li}},\ }\href {\doibase 10.1103/PhysRevD.104.083546} {\bibfield  {journal} {\bibinfo  {journal} {Phys. Rev. D}\ }\textbf {\bibinfo {volume} {104}},\ \bibinfo {pages} {083546} (\bibinfo {year} {2021})},\ \Eprint {http://arxiv.org/abs/2111.10028} {arXiv:2111.10028 [astro-ph.CO]} \BibitemShut {NoStop}%
\bibitem [{\citenamefont {Rezazadeh}\ \emph {et~al.}(2022)\citenamefont {Rezazadeh}, \citenamefont {Teimoori}, \citenamefont {Karimi},\ and\ \citenamefont {Karami}}]{Rezazadeh:2021clf}%
  \BibitemOpen
  \bibfield  {author} {\bibinfo {author} {\bibfnamefont {K.}~\bibnamefont {Rezazadeh}}, \bibinfo {author} {\bibfnamefont {Z.}~\bibnamefont {Teimoori}}, \bibinfo {author} {\bibfnamefont {S.}~\bibnamefont {Karimi}}, \ and\ \bibinfo {author} {\bibfnamefont {K.}~\bibnamefont {Karami}},\ }\href {\doibase 10.1140/epjc/s10052-022-10735-w} {\bibfield  {journal} {\bibinfo  {journal} {Eur. Phys. J. C}\ }\textbf {\bibinfo {volume} {82}},\ \bibinfo {pages} {758} (\bibinfo {year} {2022})},\ \Eprint {http://arxiv.org/abs/2110.01482} {arXiv:2110.01482 [gr-qc]} \BibitemShut {NoStop}%
\bibitem [{\citenamefont {Iacconi}\ \emph {et~al.}(2022)\citenamefont {Iacconi}, \citenamefont {Assadullahi}, \citenamefont {Fasiello},\ and\ \citenamefont {Wands}}]{Iacconi:2021ltm}%
  \BibitemOpen
  \bibfield  {author} {\bibinfo {author} {\bibfnamefont {L.}~\bibnamefont {Iacconi}}, \bibinfo {author} {\bibfnamefont {H.}~\bibnamefont {Assadullahi}}, \bibinfo {author} {\bibfnamefont {M.}~\bibnamefont {Fasiello}}, \ and\ \bibinfo {author} {\bibfnamefont {D.}~\bibnamefont {Wands}},\ }\href {\doibase 10.1088/1475-7516/2022/06/007} {\bibfield  {journal} {\bibinfo  {journal} {JCAP}\ }\textbf {\bibinfo {volume} {06}},\ \bibinfo {pages} {007} (\bibinfo {year} {2022})},\ \Eprint {http://arxiv.org/abs/2112.05092} {arXiv:2112.05092 [astro-ph.CO]} \BibitemShut {NoStop}%
\bibitem [{\citenamefont {Cai}\ \emph {et~al.}(2022)\citenamefont {Cai}, \citenamefont {Ma}, \citenamefont {Sasaki}, \citenamefont {Wang},\ and\ \citenamefont {Zhou}}]{Cai:2021zsp}%
  \BibitemOpen
  \bibfield  {author} {\bibinfo {author} {\bibfnamefont {Y.-F.}\ \bibnamefont {Cai}}, \bibinfo {author} {\bibfnamefont {X.-H.}\ \bibnamefont {Ma}}, \bibinfo {author} {\bibfnamefont {M.}~\bibnamefont {Sasaki}}, \bibinfo {author} {\bibfnamefont {D.-G.}\ \bibnamefont {Wang}}, \ and\ \bibinfo {author} {\bibfnamefont {Z.}~\bibnamefont {Zhou}},\ }\href {\doibase 10.1016/j.physletb.2022.137461} {\bibfield  {journal} {\bibinfo  {journal} {Phys. Lett. B}\ }\textbf {\bibinfo {volume} {834}},\ \bibinfo {pages} {137461} (\bibinfo {year} {2022})},\ \Eprint {http://arxiv.org/abs/2112.13836} {arXiv:2112.13836 [astro-ph.CO]} \BibitemShut {NoStop}%
\bibitem [{\citenamefont {Kefala}\ \emph {et~al.}(2021)\citenamefont {Kefala}, \citenamefont {Kodaxis}, \citenamefont {Stamou},\ and\ \citenamefont {Tetradis}}]{Kefala:2020xsx}%
  \BibitemOpen
  \bibfield  {author} {\bibinfo {author} {\bibfnamefont {K.}~\bibnamefont {Kefala}}, \bibinfo {author} {\bibfnamefont {G.~P.}\ \bibnamefont {Kodaxis}}, \bibinfo {author} {\bibfnamefont {I.~D.}\ \bibnamefont {Stamou}}, \ and\ \bibinfo {author} {\bibfnamefont {N.}~\bibnamefont {Tetradis}},\ }\href {\doibase 10.1103/PhysRevD.104.023506} {\bibfield  {journal} {\bibinfo  {journal} {Phys. Rev. D}\ }\textbf {\bibinfo {volume} {104}},\ \bibinfo {pages} {023506} (\bibinfo {year} {2021})},\ \Eprint {http://arxiv.org/abs/2010.12483} {arXiv:2010.12483 [astro-ph.CO]} \BibitemShut {NoStop}%
\bibitem [{\citenamefont {Inomata}\ \emph {et~al.}(2021)\citenamefont {Inomata}, \citenamefont {McDonough},\ and\ \citenamefont {Hu}}]{Inomata:2021uqj}%
  \BibitemOpen
  \bibfield  {author} {\bibinfo {author} {\bibfnamefont {K.}~\bibnamefont {Inomata}}, \bibinfo {author} {\bibfnamefont {E.}~\bibnamefont {McDonough}}, \ and\ \bibinfo {author} {\bibfnamefont {W.}~\bibnamefont {Hu}},\ }\href {\doibase 10.1103/PhysRevD.104.123553} {\bibfield  {journal} {\bibinfo  {journal} {Phys. Rev. D}\ }\textbf {\bibinfo {volume} {104}},\ \bibinfo {pages} {123553} (\bibinfo {year} {2021})},\ \Eprint {http://arxiv.org/abs/2104.03972} {arXiv:2104.03972 [astro-ph.CO]} \BibitemShut {NoStop}%
\bibitem [{\citenamefont {Ng}\ and\ \citenamefont {Wu}(2021)}]{Ng:2021hll}%
  \BibitemOpen
  \bibfield  {author} {\bibinfo {author} {\bibfnamefont {K.-W.}\ \bibnamefont {Ng}}\ and\ \bibinfo {author} {\bibfnamefont {Y.-P.}\ \bibnamefont {Wu}},\ }\href {\doibase 10.1007/JHEP11(2021)076} {\bibfield  {journal} {\bibinfo  {journal} {JHEP}\ }\textbf {\bibinfo {volume} {11}},\ \bibinfo {pages} {076} (\bibinfo {year} {2021})},\ \Eprint {http://arxiv.org/abs/2102.05620} {arXiv:2102.05620 [astro-ph.CO]} \BibitemShut {NoStop}%
\bibitem [{\citenamefont {Motohashi}\ \emph {et~al.}(2020)\citenamefont {Motohashi}, \citenamefont {Mukohyama},\ and\ \citenamefont {Oliosi}}]{Motohashi:2019rhu}%
  \BibitemOpen
  \bibfield  {author} {\bibinfo {author} {\bibfnamefont {H.}~\bibnamefont {Motohashi}}, \bibinfo {author} {\bibfnamefont {S.}~\bibnamefont {Mukohyama}}, \ and\ \bibinfo {author} {\bibfnamefont {M.}~\bibnamefont {Oliosi}},\ }\href {\doibase 10.1088/1475-7516/2020/03/002} {\bibfield  {journal} {\bibinfo  {journal} {JCAP}\ }\textbf {\bibinfo {volume} {03}},\ \bibinfo {pages} {002} (\bibinfo {year} {2020})},\ \Eprint {http://arxiv.org/abs/1910.13235} {arXiv:1910.13235 [gr-qc]} \BibitemShut {NoStop}%
\bibitem [{\citenamefont {Inomata}\ \emph {et~al.}(2022)\citenamefont {Inomata}, \citenamefont {McDonough},\ and\ \citenamefont {Hu}}]{Inomata:2021tpx}%
  \BibitemOpen
  \bibfield  {author} {\bibinfo {author} {\bibfnamefont {K.}~\bibnamefont {Inomata}}, \bibinfo {author} {\bibfnamefont {E.}~\bibnamefont {McDonough}}, \ and\ \bibinfo {author} {\bibfnamefont {W.}~\bibnamefont {Hu}},\ }\href {\doibase 10.1088/1475-7516/2022/02/031} {\bibfield  {journal} {\bibinfo  {journal} {JCAP}\ }\textbf {\bibinfo {volume} {02}},\ \bibinfo {pages} {031} (\bibinfo {year} {2022})},\ \Eprint {http://arxiv.org/abs/2110.14641} {arXiv:2110.14641 [astro-ph.CO]} \BibitemShut {NoStop}%
\bibitem [{\citenamefont {Hertzberg}\ and\ \citenamefont {Yamada}(2018)}]{Hertzberg:2017dkh}%
  \BibitemOpen
  \bibfield  {author} {\bibinfo {author} {\bibfnamefont {M.~P.}\ \bibnamefont {Hertzberg}}\ and\ \bibinfo {author} {\bibfnamefont {M.}~\bibnamefont {Yamada}},\ }\href {\doibase 10.1103/PhysRevD.97.083509} {\bibfield  {journal} {\bibinfo  {journal} {Phys. Rev. D}\ }\textbf {\bibinfo {volume} {97}},\ \bibinfo {pages} {083509} (\bibinfo {year} {2018})},\ \Eprint {http://arxiv.org/abs/1712.09750} {arXiv:1712.09750 [astro-ph.CO]} \BibitemShut {NoStop}%
\bibitem [{\citenamefont {Ballesteros}\ \emph {et~al.}(2020{\natexlab{b}})\citenamefont {Ballesteros}, \citenamefont {Rey}, \citenamefont {Taoso},\ and\ \citenamefont {Urbano}}]{Ballesteros:2020qam}%
  \BibitemOpen
  \bibfield  {author} {\bibinfo {author} {\bibfnamefont {G.}~\bibnamefont {Ballesteros}}, \bibinfo {author} {\bibfnamefont {J.}~\bibnamefont {Rey}}, \bibinfo {author} {\bibfnamefont {M.}~\bibnamefont {Taoso}}, \ and\ \bibinfo {author} {\bibfnamefont {A.}~\bibnamefont {Urbano}},\ }\href {\doibase 10.1088/1475-7516/2020/07/025} {\bibfield  {journal} {\bibinfo  {journal} {JCAP}\ }\textbf {\bibinfo {volume} {07}},\ \bibinfo {pages} {025} (\bibinfo {year} {2020}{\natexlab{b}})},\ \Eprint {http://arxiv.org/abs/2001.08220} {arXiv:2001.08220 [astro-ph.CO]} \BibitemShut {NoStop}%
\bibitem [{\citenamefont {Kannike}\ \emph {et~al.}(2017)\citenamefont {Kannike}, \citenamefont {Marzola}, \citenamefont {Raidal},\ and\ \citenamefont {Veerm\"ae}}]{Kannike:2017bxn}%
  \BibitemOpen
  \bibfield  {author} {\bibinfo {author} {\bibfnamefont {K.}~\bibnamefont {Kannike}}, \bibinfo {author} {\bibfnamefont {L.}~\bibnamefont {Marzola}}, \bibinfo {author} {\bibfnamefont {M.}~\bibnamefont {Raidal}}, \ and\ \bibinfo {author} {\bibfnamefont {H.}~\bibnamefont {Veerm\"ae}},\ }\href {\doibase 10.1088/1475-7516/2017/09/020} {\bibfield  {journal} {\bibinfo  {journal} {JCAP}\ }\textbf {\bibinfo {volume} {09}},\ \bibinfo {pages} {020} (\bibinfo {year} {2017})},\ \Eprint {http://arxiv.org/abs/1705.06225} {arXiv:1705.06225 [astro-ph.CO]} \BibitemShut {NoStop}%
\bibitem [{\citenamefont {Di}\ and\ \citenamefont {Gong}(2018)}]{Di:2017ndc}%
  \BibitemOpen
  \bibfield  {author} {\bibinfo {author} {\bibfnamefont {H.}~\bibnamefont {Di}}\ and\ \bibinfo {author} {\bibfnamefont {Y.}~\bibnamefont {Gong}},\ }\href {\doibase 10.1088/1475-7516/2018/07/007} {\bibfield  {journal} {\bibinfo  {journal} {JCAP}\ }\textbf {\bibinfo {volume} {07}},\ \bibinfo {pages} {007} (\bibinfo {year} {2018})},\ \Eprint {http://arxiv.org/abs/1707.09578} {arXiv:1707.09578 [astro-ph.CO]} \BibitemShut {NoStop}%
\bibitem [{\citenamefont {Saito}\ \emph {et~al.}(2008)\citenamefont {Saito}, \citenamefont {Yokoyama},\ and\ \citenamefont {Nagata}}]{Saito:2008em}%
  \BibitemOpen
  \bibfield  {author} {\bibinfo {author} {\bibfnamefont {R.}~\bibnamefont {Saito}}, \bibinfo {author} {\bibfnamefont {J.}~\bibnamefont {Yokoyama}}, \ and\ \bibinfo {author} {\bibfnamefont {R.}~\bibnamefont {Nagata}},\ }\href {\doibase 10.1088/1475-7516/2008/06/024} {\bibfield  {journal} {\bibinfo  {journal} {JCAP}\ }\textbf {\bibinfo {volume} {06}},\ \bibinfo {pages} {024} (\bibinfo {year} {2008})},\ \Eprint {http://arxiv.org/abs/0804.3470} {arXiv:0804.3470 [astro-ph]} \BibitemShut {NoStop}%
\bibitem [{\citenamefont {Bugaev}\ and\ \citenamefont {Klimai}(2008)}]{Bugaev:2008bi}%
  \BibitemOpen
  \bibfield  {author} {\bibinfo {author} {\bibfnamefont {E.}~\bibnamefont {Bugaev}}\ and\ \bibinfo {author} {\bibfnamefont {P.}~\bibnamefont {Klimai}},\ }\href {\doibase 10.1103/PhysRevD.78.063515} {\bibfield  {journal} {\bibinfo  {journal} {Phys. Rev. D}\ }\textbf {\bibinfo {volume} {78}},\ \bibinfo {pages} {063515} (\bibinfo {year} {2008})},\ \Eprint {http://arxiv.org/abs/0806.4541} {arXiv:0806.4541 [astro-ph]} \BibitemShut {NoStop}%
\bibitem [{\citenamefont {Solbi}\ and\ \citenamefont {Karami}(2021{\natexlab{a}})}]{Solbi:2021wbo}%
  \BibitemOpen
  \bibfield  {author} {\bibinfo {author} {\bibfnamefont {M.}~\bibnamefont {Solbi}}\ and\ \bibinfo {author} {\bibfnamefont {K.}~\bibnamefont {Karami}},\ }\href {\doibase 10.1088/1475-7516/2021/08/056} {\bibfield  {journal} {\bibinfo  {journal} {JCAP}\ }\textbf {\bibinfo {volume} {08}},\ \bibinfo {pages} {056} (\bibinfo {year} {2021}{\natexlab{a}})},\ \Eprint {http://arxiv.org/abs/2102.05651} {arXiv:2102.05651 [astro-ph.CO]} \BibitemShut {NoStop}%
\bibitem [{\citenamefont {Solbi}\ and\ \citenamefont {Karami}(2021{\natexlab{b}})}]{Solbi:2021rse}%
  \BibitemOpen
  \bibfield  {author} {\bibinfo {author} {\bibfnamefont {M.}~\bibnamefont {Solbi}}\ and\ \bibinfo {author} {\bibfnamefont {K.}~\bibnamefont {Karami}},\ }\href {\doibase 10.1140/epjc/s10052-021-09690-9} {\bibfield  {journal} {\bibinfo  {journal} {Eur. Phys. J. C}\ }\textbf {\bibinfo {volume} {81}},\ \bibinfo {pages} {884} (\bibinfo {year} {2021}{\natexlab{b}})},\ \Eprint {http://arxiv.org/abs/2106.02863} {arXiv:2106.02863 [astro-ph.CO]} \BibitemShut {NoStop}%
\bibitem [{\citenamefont {Teimoori}\ \emph {et~al.}(2021)\citenamefont {Teimoori}, \citenamefont {Rezazadeh}, \citenamefont {Rasheed},\ and\ \citenamefont {Karami}}]{Teimoori:2021pte}%
  \BibitemOpen
  \bibfield  {author} {\bibinfo {author} {\bibfnamefont {Z.}~\bibnamefont {Teimoori}}, \bibinfo {author} {\bibfnamefont {K.}~\bibnamefont {Rezazadeh}}, \bibinfo {author} {\bibfnamefont {M.~A.}\ \bibnamefont {Rasheed}}, \ and\ \bibinfo {author} {\bibfnamefont {K.}~\bibnamefont {Karami}},\ }\href {\doibase 10.1088/1475-7516/2021/10/018} {\bibfield  {journal} {\bibinfo  {journal} {JCAP}\ }\textbf {\bibinfo {volume} {2021}},\ \bibinfo {pages} {018} (\bibinfo {year} {2021})},\ \Eprint {http://arxiv.org/abs/2107.07620} {arXiv:2107.07620 [astro-ph.CO]} \BibitemShut {NoStop}%
\bibitem [{\citenamefont {Ballesteros}\ \emph {et~al.}(2019)\citenamefont {Ballesteros}, \citenamefont {Beltran~Jimenez},\ and\ \citenamefont {Pieroni}}]{Ballesteros:2018wlw}%
  \BibitemOpen
  \bibfield  {author} {\bibinfo {author} {\bibfnamefont {G.}~\bibnamefont {Ballesteros}}, \bibinfo {author} {\bibfnamefont {J.}~\bibnamefont {Beltran~Jimenez}}, \ and\ \bibinfo {author} {\bibfnamefont {M.}~\bibnamefont {Pieroni}},\ }\href {\doibase 10.1088/1475-7516/2019/06/016} {\bibfield  {journal} {\bibinfo  {journal} {JCAP}\ }\textbf {\bibinfo {volume} {06}},\ \bibinfo {pages} {016} (\bibinfo {year} {2019})},\ \Eprint {http://arxiv.org/abs/1811.03065} {arXiv:1811.03065 [astro-ph.CO]} \BibitemShut {NoStop}%
\bibitem [{\citenamefont {Ballesteros}\ \emph {et~al.}(2022)\citenamefont {Ballesteros}, \citenamefont {C\'espedes},\ and\ \citenamefont {Santoni}}]{Ballesteros:2021fsp}%
  \BibitemOpen
  \bibfield  {author} {\bibinfo {author} {\bibfnamefont {G.}~\bibnamefont {Ballesteros}}, \bibinfo {author} {\bibfnamefont {S.}~\bibnamefont {C\'espedes}}, \ and\ \bibinfo {author} {\bibfnamefont {L.}~\bibnamefont {Santoni}},\ }\href {\doibase 10.1007/JHEP01(2022)074} {\bibfield  {journal} {\bibinfo  {journal} {JHEP}\ }\textbf {\bibinfo {volume} {01}},\ \bibinfo {pages} {074} (\bibinfo {year} {2022})},\ \Eprint {http://arxiv.org/abs/2109.00567} {arXiv:2109.00567 [hep-th]} \BibitemShut {NoStop}%
\bibitem [{\citenamefont {Ashoorioon}\ \emph {et~al.}(2021)\citenamefont {Ashoorioon}, \citenamefont {Rostami},\ and\ \citenamefont {Firouzjaee}}]{Ashoorioon:2019xqc}%
  \BibitemOpen
  \bibfield  {author} {\bibinfo {author} {\bibfnamefont {A.}~\bibnamefont {Ashoorioon}}, \bibinfo {author} {\bibfnamefont {A.}~\bibnamefont {Rostami}}, \ and\ \bibinfo {author} {\bibfnamefont {J.~T.}\ \bibnamefont {Firouzjaee}},\ }\href {\doibase 10.1007/JHEP07(2021)087} {\bibfield  {journal} {\bibinfo  {journal} {JHEP}\ }\textbf {\bibinfo {volume} {07}},\ \bibinfo {pages} {087} (\bibinfo {year} {2021})},\ \Eprint {http://arxiv.org/abs/1912.13326} {arXiv:1912.13326 [astro-ph.CO]} \BibitemShut {NoStop}%
\bibitem [{\citenamefont {Frolovsky}\ \emph {et~al.}(2022)\citenamefont {Frolovsky}, \citenamefont {Ketov},\ and\ \citenamefont {Saburov}}]{Frolovsky:2022ewg}%
  \BibitemOpen
  \bibfield  {author} {\bibinfo {author} {\bibfnamefont {D.}~\bibnamefont {Frolovsky}}, \bibinfo {author} {\bibfnamefont {S.~V.}\ \bibnamefont {Ketov}}, \ and\ \bibinfo {author} {\bibfnamefont {S.}~\bibnamefont {Saburov}},\ }\href {\doibase 10.1142/S0217732322501358} {\bibfield  {journal} {\bibinfo  {journal} {Mod. Phys. Lett. A}\ }\textbf {\bibinfo {volume} {37}},\ \bibinfo {pages} {2250135} (\bibinfo {year} {2022})},\ \Eprint {http://arxiv.org/abs/2205.00603} {arXiv:2205.00603 [astro-ph.CO]} \BibitemShut {NoStop}%
\bibitem [{\citenamefont {Fu}\ \emph {et~al.}(2019)\citenamefont {Fu}, \citenamefont {Wu},\ and\ \citenamefont {Yu}}]{Fu:2019ttf}%
  \BibitemOpen
  \bibfield  {author} {\bibinfo {author} {\bibfnamefont {C.}~\bibnamefont {Fu}}, \bibinfo {author} {\bibfnamefont {P.}~\bibnamefont {Wu}}, \ and\ \bibinfo {author} {\bibfnamefont {H.}~\bibnamefont {Yu}},\ }\href {\doibase 10.1103/PhysRevD.100.063532} {\bibfield  {journal} {\bibinfo  {journal} {Phys. Rev. D}\ }\textbf {\bibinfo {volume} {100}},\ \bibinfo {pages} {063532} (\bibinfo {year} {2019})},\ \Eprint {http://arxiv.org/abs/1907.05042} {arXiv:1907.05042 [astro-ph.CO]} \BibitemShut {NoStop}%
\bibitem [{\citenamefont {Heydari}\ and\ \citenamefont {Karami}(2022)}]{Heydari:2021gea}%
  \BibitemOpen
  \bibfield  {author} {\bibinfo {author} {\bibfnamefont {S.}~\bibnamefont {Heydari}}\ and\ \bibinfo {author} {\bibfnamefont {K.}~\bibnamefont {Karami}},\ }\href {\doibase 10.1140/epjc/s10052-022-10036-2} {\bibfield  {journal} {\bibinfo  {journal} {Eur. Phys. J. C}\ }\textbf {\bibinfo {volume} {82}},\ \bibinfo {pages} {83} (\bibinfo {year} {2022})},\ \Eprint {http://arxiv.org/abs/2107.10550} {arXiv:2107.10550 [gr-qc]} \BibitemShut {NoStop}%
\bibitem [{\citenamefont {Kawai}\ and\ \citenamefont {Kim}(2021)}]{Kawai:2021edk}%
  \BibitemOpen
  \bibfield  {author} {\bibinfo {author} {\bibfnamefont {S.}~\bibnamefont {Kawai}}\ and\ \bibinfo {author} {\bibfnamefont {J.}~\bibnamefont {Kim}},\ }\href {\doibase 10.1103/PhysRevD.104.083545} {\bibfield  {journal} {\bibinfo  {journal} {Phys. Rev. D}\ }\textbf {\bibinfo {volume} {104}},\ \bibinfo {pages} {083545} (\bibinfo {year} {2021})},\ \Eprint {http://arxiv.org/abs/2108.01340} {arXiv:2108.01340 [astro-ph.CO]} \BibitemShut {NoStop}%
\bibitem [{\citenamefont {\"Ozsoy}\ and\ \citenamefont {Lalak}(2021)}]{Ozsoy:2020kat}%
  \BibitemOpen
  \bibfield  {author} {\bibinfo {author} {\bibfnamefont {O.}~\bibnamefont {\"Ozsoy}}\ and\ \bibinfo {author} {\bibfnamefont {Z.}~\bibnamefont {Lalak}},\ }\href {\doibase 10.1088/1475-7516/2021/01/040} {\bibfield  {journal} {\bibinfo  {journal} {JCAP}\ }\textbf {\bibinfo {volume} {01}},\ \bibinfo {pages} {040} (\bibinfo {year} {2021})},\ \Eprint {http://arxiv.org/abs/2008.07549} {arXiv:2008.07549 [astro-ph.CO]} \BibitemShut {NoStop}%
\bibitem [{\citenamefont {Carr}\ \emph {et~al.}(2010)\citenamefont {Carr}, \citenamefont {Kohri}, \citenamefont {Sendouda},\ and\ \citenamefont {Yokoyama}}]{Carr:2009jm}%
  \BibitemOpen
  \bibfield  {author} {\bibinfo {author} {\bibfnamefont {B.~J.}\ \bibnamefont {Carr}}, \bibinfo {author} {\bibfnamefont {K.}~\bibnamefont {Kohri}}, \bibinfo {author} {\bibfnamefont {Y.}~\bibnamefont {Sendouda}}, \ and\ \bibinfo {author} {\bibfnamefont {J.}~\bibnamefont {Yokoyama}},\ }\href {\doibase 10.1103/PhysRevD.81.104019} {\bibfield  {journal} {\bibinfo  {journal} {Phys. Rev. D}\ }\textbf {\bibinfo {volume} {81}},\ \bibinfo {pages} {104019} (\bibinfo {year} {2010})},\ \Eprint {http://arxiv.org/abs/0912.5297} {arXiv:0912.5297 [astro-ph.CO]} \BibitemShut {NoStop}%
\bibitem [{\citenamefont {Carr}\ \emph {et~al.}(2021)\citenamefont {Carr}, \citenamefont {Kohri}, \citenamefont {Sendouda},\ and\ \citenamefont {Yokoyama}}]{Carr:2020gox}%
  \BibitemOpen
  \bibfield  {author} {\bibinfo {author} {\bibfnamefont {B.}~\bibnamefont {Carr}}, \bibinfo {author} {\bibfnamefont {K.}~\bibnamefont {Kohri}}, \bibinfo {author} {\bibfnamefont {Y.}~\bibnamefont {Sendouda}}, \ and\ \bibinfo {author} {\bibfnamefont {J.}~\bibnamefont {Yokoyama}},\ }\href {\doibase 10.1088/1361-6633/ac1e31} {\bibfield  {journal} {\bibinfo  {journal} {Rept. Prog. Phys.}\ }\textbf {\bibinfo {volume} {84}},\ \bibinfo {pages} {116902} (\bibinfo {year} {2021})},\ \Eprint {http://arxiv.org/abs/2002.12778} {arXiv:2002.12778 [astro-ph.CO]} \BibitemShut {NoStop}%
\bibitem [{\citenamefont {Assadullahi}\ and\ \citenamefont {Wands}(2010)}]{Assadullahi:2009jc}%
  \BibitemOpen
  \bibfield  {author} {\bibinfo {author} {\bibfnamefont {H.}~\bibnamefont {Assadullahi}}\ and\ \bibinfo {author} {\bibfnamefont {D.}~\bibnamefont {Wands}},\ }\href {\doibase 10.1103/PhysRevD.81.023527} {\bibfield  {journal} {\bibinfo  {journal} {Phys. Rev. D}\ }\textbf {\bibinfo {volume} {81}},\ \bibinfo {pages} {023527} (\bibinfo {year} {2010})},\ \Eprint {http://arxiv.org/abs/0907.4073} {arXiv:0907.4073 [astro-ph.CO]} \BibitemShut {NoStop}%
\bibitem [{\citenamefont {Baumann}\ \emph {et~al.}(2007)\citenamefont {Baumann}, \citenamefont {Steinhardt}, \citenamefont {Takahashi},\ and\ \citenamefont {Ichiki}}]{Baumann:2007zm}%
  \BibitemOpen
  \bibfield  {author} {\bibinfo {author} {\bibfnamefont {D.}~\bibnamefont {Baumann}}, \bibinfo {author} {\bibfnamefont {P.~J.}\ \bibnamefont {Steinhardt}}, \bibinfo {author} {\bibfnamefont {K.}~\bibnamefont {Takahashi}}, \ and\ \bibinfo {author} {\bibfnamefont {K.}~\bibnamefont {Ichiki}},\ }\href {\doibase 10.1103/PhysRevD.76.084019} {\bibfield  {journal} {\bibinfo  {journal} {Phys. Rev. D}\ }\textbf {\bibinfo {volume} {76}},\ \bibinfo {pages} {084019} (\bibinfo {year} {2007})},\ \Eprint {http://arxiv.org/abs/hep-th/0703290} {arXiv:hep-th/0703290} \BibitemShut {NoStop}%
\bibitem [{\citenamefont {Saito}\ and\ \citenamefont {Yokoyama}(2009)}]{Saito:2008jc}%
  \BibitemOpen
  \bibfield  {author} {\bibinfo {author} {\bibfnamefont {R.}~\bibnamefont {Saito}}\ and\ \bibinfo {author} {\bibfnamefont {J.}~\bibnamefont {Yokoyama}},\ }\href {\doibase 10.1103/PhysRevLett.102.161101} {\bibfield  {journal} {\bibinfo  {journal} {Phys. Rev. Lett.}\ }\textbf {\bibinfo {volume} {102}},\ \bibinfo {pages} {161101} (\bibinfo {year} {2009})},\ \bibinfo {note} {[Erratum: Phys.Rev.Lett. 107, 069901 (2011)]},\ \Eprint {http://arxiv.org/abs/0812.4339} {arXiv:0812.4339 [astro-ph]} \BibitemShut {NoStop}%
\bibitem [{\citenamefont {Saito}\ and\ \citenamefont {Yokoyama}(2010)}]{Saito:2009jt}%
  \BibitemOpen
  \bibfield  {author} {\bibinfo {author} {\bibfnamefont {R.}~\bibnamefont {Saito}}\ and\ \bibinfo {author} {\bibfnamefont {J.}~\bibnamefont {Yokoyama}},\ }\href {\doibase 10.1143/PTP.126.351} {\bibfield  {journal} {\bibinfo  {journal} {Prog. Theor. Phys.}\ }\textbf {\bibinfo {volume} {123}},\ \bibinfo {pages} {867} (\bibinfo {year} {2010})},\ \bibinfo {note} {[Erratum: Prog.Theor.Phys. 126, 351--352 (2011)]},\ \Eprint {http://arxiv.org/abs/0912.5317} {arXiv:0912.5317 [astro-ph.CO]} \BibitemShut {NoStop}%
\bibitem [{\citenamefont {Cai}\ \emph {et~al.}(2019)\citenamefont {Cai}, \citenamefont {Pi},\ and\ \citenamefont {Sasaki}}]{Cai:2018dig}%
  \BibitemOpen
  \bibfield  {author} {\bibinfo {author} {\bibfnamefont {R.-g.}\ \bibnamefont {Cai}}, \bibinfo {author} {\bibfnamefont {S.}~\bibnamefont {Pi}}, \ and\ \bibinfo {author} {\bibfnamefont {M.}~\bibnamefont {Sasaki}},\ }\href {\doibase 10.1103/PhysRevLett.122.201101} {\bibfield  {journal} {\bibinfo  {journal} {Phys. Rev. Lett.}\ }\textbf {\bibinfo {volume} {122}},\ \bibinfo {pages} {201101} (\bibinfo {year} {2019})},\ \Eprint {http://arxiv.org/abs/1810.11000} {arXiv:1810.11000 [astro-ph.CO]} \BibitemShut {NoStop}%
\bibitem [{\citenamefont {Yokoyama}(2021)}]{Yokoyama:2021hsa}%
  \BibitemOpen
  \bibfield  {author} {\bibinfo {author} {\bibfnamefont {J.}~\bibnamefont {Yokoyama}},\ }\href {\doibase 10.1007/s43673-021-00020-5} {\bibfield  {journal} {\bibinfo  {journal} {AAPPS Bull.}\ }\textbf {\bibinfo {volume} {31}},\ \bibinfo {pages} {17} (\bibinfo {year} {2021})},\ \Eprint {http://arxiv.org/abs/2105.07629} {arXiv:2105.07629 [gr-qc]} \BibitemShut {NoStop}%
\bibitem [{\citenamefont {Sato}\ and\ \citenamefont {Yokoyama}(2015)}]{Sato:2015dga}%
  \BibitemOpen
  \bibfield  {author} {\bibinfo {author} {\bibfnamefont {K.}~\bibnamefont {Sato}}\ and\ \bibinfo {author} {\bibfnamefont {J.}~\bibnamefont {Yokoyama}},\ }\href {\doibase 10.1142/S0218271815300256} {\bibfield  {journal} {\bibinfo  {journal} {Int. J. Mod. Phys. D}\ }\textbf {\bibinfo {volume} {24}},\ \bibinfo {pages} {1530025} (\bibinfo {year} {2015})}\BibitemShut {NoStop}%
\bibitem [{\citenamefont {Polarski}\ and\ \citenamefont {Starobinsky}(1996)}]{Polarski:1995jg}%
  \BibitemOpen
  \bibfield  {author} {\bibinfo {author} {\bibfnamefont {D.}~\bibnamefont {Polarski}}\ and\ \bibinfo {author} {\bibfnamefont {A.~A.}\ \bibnamefont {Starobinsky}},\ }\href {\doibase 10.1088/0264-9381/13/3/006} {\bibfield  {journal} {\bibinfo  {journal} {Class. Quant. Grav.}\ }\textbf {\bibinfo {volume} {13}},\ \bibinfo {pages} {377} (\bibinfo {year} {1996})},\ \Eprint {http://arxiv.org/abs/gr-qc/9504030} {arXiv:gr-qc/9504030} \BibitemShut {NoStop}%
\bibitem [{\citenamefont {Kinney}(1997)}]{Kinney:1997ne}%
  \BibitemOpen
  \bibfield  {author} {\bibinfo {author} {\bibfnamefont {W.~H.}\ \bibnamefont {Kinney}},\ }\href {\doibase 10.1103/PhysRevD.56.2002} {\bibfield  {journal} {\bibinfo  {journal} {Phys. Rev. D}\ }\textbf {\bibinfo {volume} {56}},\ \bibinfo {pages} {2002} (\bibinfo {year} {1997})},\ \Eprint {http://arxiv.org/abs/hep-ph/9702427} {arXiv:hep-ph/9702427} \BibitemShut {NoStop}%
\bibitem [{\citenamefont {Yokoyama}\ and\ \citenamefont {Inoue}(2002)}]{Inoue:2001zt}%
  \BibitemOpen
  \bibfield  {author} {\bibinfo {author} {\bibfnamefont {J.}~\bibnamefont {Yokoyama}}\ and\ \bibinfo {author} {\bibfnamefont {S.}~\bibnamefont {Inoue}},\ }\href {\doibase 10.1016/S0370-2693(01)01369-7} {\bibfield  {journal} {\bibinfo  {journal} {Phys. Lett. B}\ }\textbf {\bibinfo {volume} {524}},\ \bibinfo {pages} {15} (\bibinfo {year} {2002})},\ \Eprint {http://arxiv.org/abs/hep-ph/0104083} {arXiv:hep-ph/0104083} \BibitemShut {NoStop}%
\bibitem [{\citenamefont {Kinney}(2005)}]{Kinney:2005vj}%
  \BibitemOpen
  \bibfield  {author} {\bibinfo {author} {\bibfnamefont {W.~H.}\ \bibnamefont {Kinney}},\ }\href {\doibase 10.1103/PhysRevD.72.023515} {\bibfield  {journal} {\bibinfo  {journal} {Phys. Rev. D}\ }\textbf {\bibinfo {volume} {72}},\ \bibinfo {pages} {023515} (\bibinfo {year} {2005})},\ \Eprint {http://arxiv.org/abs/gr-qc/0503017} {arXiv:gr-qc/0503017} \BibitemShut {NoStop}%
\bibitem [{\citenamefont {Martin}\ \emph {et~al.}(2013)\citenamefont {Martin}, \citenamefont {Motohashi},\ and\ \citenamefont {Suyama}}]{Martin:2012pe}%
  \BibitemOpen
  \bibfield  {author} {\bibinfo {author} {\bibfnamefont {J.}~\bibnamefont {Martin}}, \bibinfo {author} {\bibfnamefont {H.}~\bibnamefont {Motohashi}}, \ and\ \bibinfo {author} {\bibfnamefont {T.}~\bibnamefont {Suyama}},\ }\href {\doibase 10.1103/PhysRevD.87.023514} {\bibfield  {journal} {\bibinfo  {journal} {Phys. Rev. D}\ }\textbf {\bibinfo {volume} {87}},\ \bibinfo {pages} {023514} (\bibinfo {year} {2013})},\ \Eprint {http://arxiv.org/abs/1211.0083} {arXiv:1211.0083 [astro-ph.CO]} \BibitemShut {NoStop}%
\bibitem [{\citenamefont {Motohashi}\ and\ \citenamefont {Hu}(2017)}]{Motohashi:2017kbs}%
  \BibitemOpen
  \bibfield  {author} {\bibinfo {author} {\bibfnamefont {H.}~\bibnamefont {Motohashi}}\ and\ \bibinfo {author} {\bibfnamefont {W.}~\bibnamefont {Hu}},\ }\href {\doibase 10.1103/PhysRevD.96.063503} {\bibfield  {journal} {\bibinfo  {journal} {Phys. Rev. D}\ }\textbf {\bibinfo {volume} {96}},\ \bibinfo {pages} {063503} (\bibinfo {year} {2017})},\ \Eprint {http://arxiv.org/abs/1706.06784} {arXiv:1706.06784 [astro-ph.CO]} \BibitemShut {NoStop}%
\bibitem [{\citenamefont {Yokoyama}(1999)}]{Yokoyama:1998rw}%
  \BibitemOpen
  \bibfield  {author} {\bibinfo {author} {\bibfnamefont {J.}~\bibnamefont {Yokoyama}},\ }\href {\doibase 10.1103/PhysRevD.59.107303} {\bibfield  {journal} {\bibinfo  {journal} {Phys. Rev. D}\ }\textbf {\bibinfo {volume} {59}},\ \bibinfo {pages} {107303} (\bibinfo {year} {1999})}\BibitemShut {NoStop}%
\bibitem [{\citenamefont {Motohashi}\ \emph {et~al.}(2015)\citenamefont {Motohashi}, \citenamefont {Starobinsky},\ and\ \citenamefont {Yokoyama}}]{Motohashi:2014ppa}%
  \BibitemOpen
  \bibfield  {author} {\bibinfo {author} {\bibfnamefont {H.}~\bibnamefont {Motohashi}}, \bibinfo {author} {\bibfnamefont {A.~A.}\ \bibnamefont {Starobinsky}}, \ and\ \bibinfo {author} {\bibfnamefont {J.}~\bibnamefont {Yokoyama}},\ }\href {\doibase 10.1088/1475-7516/2015/09/018} {\bibfield  {journal} {\bibinfo  {journal} {JCAP}\ }\textbf {\bibinfo {volume} {09}},\ \bibinfo {pages} {018} (\bibinfo {year} {2015})},\ \Eprint {http://arxiv.org/abs/1411.5021} {arXiv:1411.5021 [astro-ph.CO]} \BibitemShut {NoStop}%
\bibitem [{\citenamefont {Maldacena}(2003)}]{Maldacena:2002vr}%
  \BibitemOpen
  \bibfield  {author} {\bibinfo {author} {\bibfnamefont {J.~M.}\ \bibnamefont {Maldacena}},\ }\href {\doibase 10.1088/1126-6708/2003/05/013} {\bibfield  {journal} {\bibinfo  {journal} {JHEP}\ }\textbf {\bibinfo {volume} {05}},\ \bibinfo {pages} {013} (\bibinfo {year} {2003})},\ \Eprint {http://arxiv.org/abs/astro-ph/0210603} {arXiv:astro-ph/0210603} \BibitemShut {NoStop}%
\bibitem [{\citenamefont {Chen}\ \emph {et~al.}(2007)\citenamefont {Chen}, \citenamefont {Huang}, \citenamefont {Kachru},\ and\ \citenamefont {Shiu}}]{Chen:2006nt}%
  \BibitemOpen
  \bibfield  {author} {\bibinfo {author} {\bibfnamefont {X.}~\bibnamefont {Chen}}, \bibinfo {author} {\bibfnamefont {M.-x.}\ \bibnamefont {Huang}}, \bibinfo {author} {\bibfnamefont {S.}~\bibnamefont {Kachru}}, \ and\ \bibinfo {author} {\bibfnamefont {G.}~\bibnamefont {Shiu}},\ }\href {\doibase 10.1088/1475-7516/2007/01/002} {\bibfield  {journal} {\bibinfo  {journal} {JCAP}\ }\textbf {\bibinfo {volume} {01}},\ \bibinfo {pages} {002} (\bibinfo {year} {2007})},\ \Eprint {http://arxiv.org/abs/hep-th/0605045} {arXiv:hep-th/0605045} \BibitemShut {NoStop}%
\bibitem [{\citenamefont {Seery}\ and\ \citenamefont {Lidsey}(2005)}]{Seery:2005wm}%
  \BibitemOpen
  \bibfield  {author} {\bibinfo {author} {\bibfnamefont {D.}~\bibnamefont {Seery}}\ and\ \bibinfo {author} {\bibfnamefont {J.~E.}\ \bibnamefont {Lidsey}},\ }\href {\doibase 10.1088/1475-7516/2005/06/003} {\bibfield  {journal} {\bibinfo  {journal} {JCAP}\ }\textbf {\bibinfo {volume} {06}},\ \bibinfo {pages} {003} (\bibinfo {year} {2005})},\ \Eprint {http://arxiv.org/abs/astro-ph/0503692} {arXiv:astro-ph/0503692} \BibitemShut {NoStop}%
\bibitem [{\citenamefont {Weinberg}(2005)}]{Weinberg:2005vy}%
  \BibitemOpen
  \bibfield  {author} {\bibinfo {author} {\bibfnamefont {S.}~\bibnamefont {Weinberg}},\ }\href {\doibase 10.1103/PhysRevD.72.043514} {\bibfield  {journal} {\bibinfo  {journal} {Phys. Rev. D}\ }\textbf {\bibinfo {volume} {72}},\ \bibinfo {pages} {043514} (\bibinfo {year} {2005})},\ \Eprint {http://arxiv.org/abs/hep-th/0506236} {arXiv:hep-th/0506236} \BibitemShut {NoStop}%
\bibitem [{\citenamefont {Sloth}(2006)}]{Sloth:2006az}%
  \BibitemOpen
  \bibfield  {author} {\bibinfo {author} {\bibfnamefont {M.~S.}\ \bibnamefont {Sloth}},\ }\href {\doibase 10.1016/j.nuclphysb.2006.04.029} {\bibfield  {journal} {\bibinfo  {journal} {Nucl. Phys. B}\ }\textbf {\bibinfo {volume} {748}},\ \bibinfo {pages} {149} (\bibinfo {year} {2006})},\ \Eprint {http://arxiv.org/abs/astro-ph/0604488} {arXiv:astro-ph/0604488} \BibitemShut {NoStop}%
\bibitem [{\citenamefont {Seery}(2008)}]{Seery:2007wf}%
  \BibitemOpen
  \bibfield  {author} {\bibinfo {author} {\bibfnamefont {D.}~\bibnamefont {Seery}},\ }\href {\doibase 10.1088/1475-7516/2008/02/006} {\bibfield  {journal} {\bibinfo  {journal} {JCAP}\ }\textbf {\bibinfo {volume} {02}},\ \bibinfo {pages} {006} (\bibinfo {year} {2008})},\ \Eprint {http://arxiv.org/abs/0707.3378} {arXiv:0707.3378 [astro-ph]} \BibitemShut {NoStop}%
\bibitem [{\citenamefont {Bartolo}\ \emph {et~al.}(2010)\citenamefont {Bartolo}, \citenamefont {Dimastrogiovanni},\ and\ \citenamefont {Vallinotto}}]{Bartolo:2010bu}%
  \BibitemOpen
  \bibfield  {author} {\bibinfo {author} {\bibfnamefont {N.}~\bibnamefont {Bartolo}}, \bibinfo {author} {\bibfnamefont {E.}~\bibnamefont {Dimastrogiovanni}}, \ and\ \bibinfo {author} {\bibfnamefont {A.}~\bibnamefont {Vallinotto}},\ }\href {\doibase 10.1088/1475-7516/2010/11/003} {\bibfield  {journal} {\bibinfo  {journal} {JCAP}\ }\textbf {\bibinfo {volume} {11}},\ \bibinfo {pages} {003} (\bibinfo {year} {2010})},\ \Eprint {http://arxiv.org/abs/1006.0196} {arXiv:1006.0196 [astro-ph.CO]} \BibitemShut {NoStop}%
\bibitem [{\citenamefont {Senatore}\ and\ \citenamefont {Zaldarriaga}(2010)}]{Senatore:2009cf}%
  \BibitemOpen
  \bibfield  {author} {\bibinfo {author} {\bibfnamefont {L.}~\bibnamefont {Senatore}}\ and\ \bibinfo {author} {\bibfnamefont {M.}~\bibnamefont {Zaldarriaga}},\ }\href {\doibase 10.1007/JHEP12(2010)008} {\bibfield  {journal} {\bibinfo  {journal} {JHEP}\ }\textbf {\bibinfo {volume} {12}},\ \bibinfo {pages} {008} (\bibinfo {year} {2010})},\ \Eprint {http://arxiv.org/abs/0912.2734} {arXiv:0912.2734 [hep-th]} \BibitemShut {NoStop}%
\bibitem [{\citenamefont {del Rio}\ \emph {et~al.}(2018)\citenamefont {del Rio}, \citenamefont {Durrer},\ and\ \citenamefont {Patil}}]{delRio:2018vrj}%
  \BibitemOpen
  \bibfield  {author} {\bibinfo {author} {\bibfnamefont {A.}~\bibnamefont {del Rio}}, \bibinfo {author} {\bibfnamefont {R.}~\bibnamefont {Durrer}}, \ and\ \bibinfo {author} {\bibfnamefont {S.~P.}\ \bibnamefont {Patil}},\ }\href {\doibase 10.1007/JHEP12(2018)094} {\bibfield  {journal} {\bibinfo  {journal} {JHEP}\ }\textbf {\bibinfo {volume} {12}},\ \bibinfo {pages} {094} (\bibinfo {year} {2018})},\ \Eprint {http://arxiv.org/abs/1808.09282} {arXiv:1808.09282 [gr-qc]} \BibitemShut {NoStop}%
\bibitem [{\citenamefont {Melville}\ and\ \citenamefont {Pajer}(2021)}]{Melville:2021lst}%
  \BibitemOpen
  \bibfield  {author} {\bibinfo {author} {\bibfnamefont {S.}~\bibnamefont {Melville}}\ and\ \bibinfo {author} {\bibfnamefont {E.}~\bibnamefont {Pajer}},\ }\href {\doibase 10.1007/JHEP05(2021)249} {\bibfield  {journal} {\bibinfo  {journal} {JHEP}\ }\textbf {\bibinfo {volume} {05}},\ \bibinfo {pages} {249} (\bibinfo {year} {2021})},\ \Eprint {http://arxiv.org/abs/2103.09832} {arXiv:2103.09832 [hep-th]} \BibitemShut {NoStop}%
\bibitem [{\citenamefont {Kristiano}\ and\ \citenamefont {Yokoyama}(2022{\natexlab{a}})}]{Kristiano:2021urj}%
  \BibitemOpen
  \bibfield  {author} {\bibinfo {author} {\bibfnamefont {J.}~\bibnamefont {Kristiano}}\ and\ \bibinfo {author} {\bibfnamefont {J.}~\bibnamefont {Yokoyama}},\ }\href {\doibase 10.1103/PhysRevLett.128.061301} {\bibfield  {journal} {\bibinfo  {journal} {Phys. Rev. Lett.}\ }\textbf {\bibinfo {volume} {128}},\ \bibinfo {pages} {061301} (\bibinfo {year} {2022}{\natexlab{a}})},\ \Eprint {http://arxiv.org/abs/2104.01953} {arXiv:2104.01953 [hep-th]} \BibitemShut {NoStop}%
\bibitem [{\citenamefont {Kristiano}\ and\ \citenamefont {Yokoyama}(2022{\natexlab{b}})}]{Kristiano:2022zpn}%
  \BibitemOpen
  \bibfield  {author} {\bibinfo {author} {\bibfnamefont {J.}~\bibnamefont {Kristiano}}\ and\ \bibinfo {author} {\bibfnamefont {J.}~\bibnamefont {Yokoyama}},\ }\href {\doibase 10.1088/1475-7516/2022/07/007} {\bibfield  {journal} {\bibinfo  {journal} {JCAP}\ }\textbf {\bibinfo {volume} {07}},\ \bibinfo {pages} {007} (\bibinfo {year} {2022}{\natexlab{b}})},\ \Eprint {http://arxiv.org/abs/2204.05202} {arXiv:2204.05202 [hep-th]} \BibitemShut {NoStop}%
\bibitem [{\citenamefont {Firouzjahi}\ \emph {et~al.}(2019)\citenamefont {Firouzjahi}, \citenamefont {Nassiri-Rad},\ and\ \citenamefont {Noorbala}}]{Firouzjahi:2018vet}%
  \BibitemOpen
  \bibfield  {author} {\bibinfo {author} {\bibfnamefont {H.}~\bibnamefont {Firouzjahi}}, \bibinfo {author} {\bibfnamefont {A.}~\bibnamefont {Nassiri-Rad}}, \ and\ \bibinfo {author} {\bibfnamefont {M.}~\bibnamefont {Noorbala}},\ }\href {\doibase 10.1088/1475-7516/2019/01/040} {\bibfield  {journal} {\bibinfo  {journal} {JCAP}\ }\textbf {\bibinfo {volume} {01}},\ \bibinfo {pages} {040} (\bibinfo {year} {2019})},\ \Eprint {http://arxiv.org/abs/1811.02175} {arXiv:1811.02175 [hep-th]} \BibitemShut {NoStop}%
\bibitem [{\citenamefont {Cheng}\ \emph {et~al.}(2022)\citenamefont {Cheng}, \citenamefont {Lee},\ and\ \citenamefont {Ng}}]{Cheng:2021lif}%
  \BibitemOpen
  \bibfield  {author} {\bibinfo {author} {\bibfnamefont {S.-L.}\ \bibnamefont {Cheng}}, \bibinfo {author} {\bibfnamefont {D.-S.}\ \bibnamefont {Lee}}, \ and\ \bibinfo {author} {\bibfnamefont {K.-W.}\ \bibnamefont {Ng}},\ }\href {\doibase 10.1016/j.physletb.2022.136956} {\bibfield  {journal} {\bibinfo  {journal} {Phys. Lett. B}\ }\textbf {\bibinfo {volume} {827}},\ \bibinfo {pages} {136956} (\bibinfo {year} {2022})},\ \Eprint {http://arxiv.org/abs/2106.09275} {arXiv:2106.09275 [astro-ph.CO]} \BibitemShut {NoStop}%
\bibitem [{\citenamefont {Syu}\ \emph {et~al.}(2020)\citenamefont {Syu}, \citenamefont {Lee},\ and\ \citenamefont {Ng}}]{Syu:2019uwx}%
  \BibitemOpen
  \bibfield  {author} {\bibinfo {author} {\bibfnamefont {W.-C.}\ \bibnamefont {Syu}}, \bibinfo {author} {\bibfnamefont {D.-S.}\ \bibnamefont {Lee}}, \ and\ \bibinfo {author} {\bibfnamefont {K.-W.}\ \bibnamefont {Ng}},\ }\href {\doibase 10.1103/PhysRevD.101.025013} {\bibfield  {journal} {\bibinfo  {journal} {Phys. Rev. D}\ }\textbf {\bibinfo {volume} {101}},\ \bibinfo {pages} {025013} (\bibinfo {year} {2020})},\ \Eprint {http://arxiv.org/abs/1907.13089} {arXiv:1907.13089 [gr-qc]} \BibitemShut {NoStop}%
\bibitem [{\citenamefont {Ando}\ and\ \citenamefont {Vennin}(2021)}]{Ando:2020fjm}%
  \BibitemOpen
  \bibfield  {author} {\bibinfo {author} {\bibfnamefont {K.}~\bibnamefont {Ando}}\ and\ \bibinfo {author} {\bibfnamefont {V.}~\bibnamefont {Vennin}},\ }\href {\doibase 10.1088/1475-7516/2021/04/057} {\bibfield  {journal} {\bibinfo  {journal} {JCAP}\ }\textbf {\bibinfo {volume} {04}},\ \bibinfo {pages} {057} (\bibinfo {year} {2021})},\ \Eprint {http://arxiv.org/abs/2012.02031} {arXiv:2012.02031 [astro-ph.CO]} \BibitemShut {NoStop}%
\bibitem [{\citenamefont {Meng}\ \emph {et~al.}(2022)\citenamefont {Meng}, \citenamefont {Yuan},\ and\ \citenamefont {Huang}}]{Meng:2022ixx}%
  \BibitemOpen
  \bibfield  {author} {\bibinfo {author} {\bibfnamefont {D.-S.}\ \bibnamefont {Meng}}, \bibinfo {author} {\bibfnamefont {C.}~\bibnamefont {Yuan}}, \ and\ \bibinfo {author} {\bibfnamefont {Q.-g.}\ \bibnamefont {Huang}},\ }\href {\doibase 10.1103/PhysRevD.106.063508} {\bibfield  {journal} {\bibinfo  {journal} {Phys. Rev. D}\ }\textbf {\bibinfo {volume} {106}},\ \bibinfo {pages} {063508} (\bibinfo {year} {2022})},\ \Eprint {http://arxiv.org/abs/2207.07668} {arXiv:2207.07668 [astro-ph.CO]} \BibitemShut {NoStop}%
\bibitem [{\citenamefont {Chen}\ \emph {et~al.}(2023)\citenamefont {Chen}, \citenamefont {Ota}, \citenamefont {Zhu},\ and\ \citenamefont {Zhu}}]{Chen:2022dah}%
  \BibitemOpen
  \bibfield  {author} {\bibinfo {author} {\bibfnamefont {C.}~\bibnamefont {Chen}}, \bibinfo {author} {\bibfnamefont {A.}~\bibnamefont {Ota}}, \bibinfo {author} {\bibfnamefont {H.-Y.}\ \bibnamefont {Zhu}}, \ and\ \bibinfo {author} {\bibfnamefont {Y.}~\bibnamefont {Zhu}},\ }\href {\doibase 10.1103/PhysRevD.107.083518} {\bibfield  {journal} {\bibinfo  {journal} {Phys. Rev. D}\ }\textbf {\bibinfo {volume} {107}},\ \bibinfo {pages} {083518} (\bibinfo {year} {2023})},\ \Eprint {http://arxiv.org/abs/2210.17176} {arXiv:2210.17176 [astro-ph.CO]} \BibitemShut {NoStop}%
\bibitem [{\citenamefont {Inomata}\ \emph {et~al.}(2023)\citenamefont {Inomata}, \citenamefont {Braglia}, \citenamefont {Chen},\ and\ \citenamefont {Renaux-Petel}}]{Inomata:2022yte}%
  \BibitemOpen
  \bibfield  {author} {\bibinfo {author} {\bibfnamefont {K.}~\bibnamefont {Inomata}}, \bibinfo {author} {\bibfnamefont {M.}~\bibnamefont {Braglia}}, \bibinfo {author} {\bibfnamefont {X.}~\bibnamefont {Chen}}, \ and\ \bibinfo {author} {\bibfnamefont {S.}~\bibnamefont {Renaux-Petel}},\ }\href {\doibase 10.1088/1475-7516/2023/04/011} {\bibfield  {journal} {\bibinfo  {journal} {JCAP}\ }\textbf {\bibinfo {volume} {04}},\ \bibinfo {pages} {011} (\bibinfo {year} {2023})},\ \Eprint {http://arxiv.org/abs/2211.02586} {arXiv:2211.02586 [astro-ph.CO]} \BibitemShut {NoStop}%
\bibitem [{\citenamefont {Sasaki}(1986)}]{Sasaki:1986hm}%
  \BibitemOpen
  \bibfield  {author} {\bibinfo {author} {\bibfnamefont {M.}~\bibnamefont {Sasaki}},\ }\href {\doibase 10.1143/PTP.76.1036} {\bibfield  {journal} {\bibinfo  {journal} {Prog. Theor. Phys.}\ }\textbf {\bibinfo {volume} {76}},\ \bibinfo {pages} {1036} (\bibinfo {year} {1986})}\BibitemShut {NoStop}%
\bibitem [{\citenamefont {Karam}\ \emph {et~al.}(2023)\citenamefont {Karam}, \citenamefont {Koivunen}, \citenamefont {Tomberg}, \citenamefont {Vaskonen},\ and\ \citenamefont {Veerm\"ae}}]{Karam:2022nym}%
  \BibitemOpen
  \bibfield  {author} {\bibinfo {author} {\bibfnamefont {A.}~\bibnamefont {Karam}}, \bibinfo {author} {\bibfnamefont {N.}~\bibnamefont {Koivunen}}, \bibinfo {author} {\bibfnamefont {E.}~\bibnamefont {Tomberg}}, \bibinfo {author} {\bibfnamefont {V.}~\bibnamefont {Vaskonen}}, \ and\ \bibinfo {author} {\bibfnamefont {H.}~\bibnamefont {Veerm\"ae}},\ }\href {\doibase 10.1088/1475-7516/2023/03/013} {\bibfield  {journal} {\bibinfo  {journal} {JCAP}\ }\textbf {\bibinfo {volume} {03}},\ \bibinfo {pages} {013} (\bibinfo {year} {2023})},\ \Eprint {http://arxiv.org/abs/2205.13540} {arXiv:2205.13540 [astro-ph.CO]} \BibitemShut {NoStop}%
\bibitem [{\citenamefont {Starobinsky}(1992)}]{Starobinsky:1992ts}%
  \BibitemOpen
  \bibfield  {author} {\bibinfo {author} {\bibfnamefont {A.~A.}\ \bibnamefont {Starobinsky}},\ }\href@noop {} {\bibfield  {journal} {\bibinfo  {journal} {JETP Lett.}\ }\textbf {\bibinfo {volume} {55}},\ \bibinfo {pages} {489} (\bibinfo {year} {1992})}\BibitemShut {NoStop}%
\bibitem [{\citenamefont {Leach}\ \emph {et~al.}(2001)\citenamefont {Leach}, \citenamefont {Sasaki}, \citenamefont {Wands},\ and\ \citenamefont {Liddle}}]{Leach:2001zf}%
  \BibitemOpen
  \bibfield  {author} {\bibinfo {author} {\bibfnamefont {S.~M.}\ \bibnamefont {Leach}}, \bibinfo {author} {\bibfnamefont {M.}~\bibnamefont {Sasaki}}, \bibinfo {author} {\bibfnamefont {D.}~\bibnamefont {Wands}}, \ and\ \bibinfo {author} {\bibfnamefont {A.~R.}\ \bibnamefont {Liddle}},\ }\href {\doibase 10.1103/PhysRevD.64.023512} {\bibfield  {journal} {\bibinfo  {journal} {Phys. Rev. D}\ }\textbf {\bibinfo {volume} {64}},\ \bibinfo {pages} {023512} (\bibinfo {year} {2001})},\ \Eprint {http://arxiv.org/abs/astro-ph/0101406} {arXiv:astro-ph/0101406} \BibitemShut {NoStop}%
\bibitem [{\citenamefont {Byrnes}\ \emph {et~al.}(2019)\citenamefont {Byrnes}, \citenamefont {Cole},\ and\ \citenamefont {Patil}}]{Byrnes:2018txb}%
  \BibitemOpen
  \bibfield  {author} {\bibinfo {author} {\bibfnamefont {C.~T.}\ \bibnamefont {Byrnes}}, \bibinfo {author} {\bibfnamefont {P.~S.}\ \bibnamefont {Cole}}, \ and\ \bibinfo {author} {\bibfnamefont {S.~P.}\ \bibnamefont {Patil}},\ }\href {\doibase 10.1088/1475-7516/2019/06/028} {\bibfield  {journal} {\bibinfo  {journal} {JCAP}\ }\textbf {\bibinfo {volume} {06}},\ \bibinfo {pages} {028} (\bibinfo {year} {2019})},\ \Eprint {http://arxiv.org/abs/1811.11158} {arXiv:1811.11158 [astro-ph.CO]} \BibitemShut {NoStop}%
\bibitem [{\citenamefont {Liu}\ \emph {et~al.}(2020)\citenamefont {Liu}, \citenamefont {Guo},\ and\ \citenamefont {Cai}}]{Liu:2020oqe}%
  \BibitemOpen
  \bibfield  {author} {\bibinfo {author} {\bibfnamefont {J.}~\bibnamefont {Liu}}, \bibinfo {author} {\bibfnamefont {Z.-K.}\ \bibnamefont {Guo}}, \ and\ \bibinfo {author} {\bibfnamefont {R.-G.}\ \bibnamefont {Cai}},\ }\href {\doibase 10.1103/PhysRevD.101.083535} {\bibfield  {journal} {\bibinfo  {journal} {Phys. Rev. D}\ }\textbf {\bibinfo {volume} {101}},\ \bibinfo {pages} {083535} (\bibinfo {year} {2020})},\ \Eprint {http://arxiv.org/abs/2003.02075} {arXiv:2003.02075 [astro-ph.CO]} \BibitemShut {NoStop}%
\bibitem [{\citenamefont {Tasinato}(2021)}]{Tasinato:2020vdk}%
  \BibitemOpen
  \bibfield  {author} {\bibinfo {author} {\bibfnamefont {G.}~\bibnamefont {Tasinato}},\ }\href {\doibase 10.1103/PhysRevD.103.023535} {\bibfield  {journal} {\bibinfo  {journal} {Phys. Rev. D}\ }\textbf {\bibinfo {volume} {103}},\ \bibinfo {pages} {023535} (\bibinfo {year} {2021})},\ \Eprint {http://arxiv.org/abs/2012.02518} {arXiv:2012.02518 [hep-th]} \BibitemShut {NoStop}%
\bibitem [{\citenamefont {Pi}\ and\ \citenamefont {Wang}(2023)}]{Pi:2022zxs}%
  \BibitemOpen
  \bibfield  {author} {\bibinfo {author} {\bibfnamefont {S.}~\bibnamefont {Pi}}\ and\ \bibinfo {author} {\bibfnamefont {J.}~\bibnamefont {Wang}},\ }\href {\doibase 10.1088/1475-7516/2023/06/018} {\bibfield  {journal} {\bibinfo  {journal} {JCAP}\ }\textbf {\bibinfo {volume} {06}},\ \bibinfo {pages} {018} (\bibinfo {year} {2023})},\ \Eprint {http://arxiv.org/abs/2209.14183} {arXiv:2209.14183 [astro-ph.CO]} \BibitemShut {NoStop}%
\bibitem [{\citenamefont {Namjoo}\ \emph {et~al.}(2013)\citenamefont {Namjoo}, \citenamefont {Firouzjahi},\ and\ \citenamefont {Sasaki}}]{Namjoo:2012aa}%
  \BibitemOpen
  \bibfield  {author} {\bibinfo {author} {\bibfnamefont {M.~H.}\ \bibnamefont {Namjoo}}, \bibinfo {author} {\bibfnamefont {H.}~\bibnamefont {Firouzjahi}}, \ and\ \bibinfo {author} {\bibfnamefont {M.}~\bibnamefont {Sasaki}},\ }\href {\doibase 10.1209/0295-5075/101/39001} {\bibfield  {journal} {\bibinfo  {journal} {EPL}\ }\textbf {\bibinfo {volume} {101}},\ \bibinfo {pages} {39001} (\bibinfo {year} {2013})},\ \Eprint {http://arxiv.org/abs/1210.3692} {arXiv:1210.3692 [astro-ph.CO]} \BibitemShut {NoStop}%
\bibitem [{\citenamefont {Cai}\ \emph {et~al.}(2016)\citenamefont {Cai}, \citenamefont {Gong}, \citenamefont {Wang},\ and\ \citenamefont {Wang}}]{Cai:2016ngx}%
  \BibitemOpen
  \bibfield  {author} {\bibinfo {author} {\bibfnamefont {Y.-F.}\ \bibnamefont {Cai}}, \bibinfo {author} {\bibfnamefont {J.-O.}\ \bibnamefont {Gong}}, \bibinfo {author} {\bibfnamefont {D.-G.}\ \bibnamefont {Wang}}, \ and\ \bibinfo {author} {\bibfnamefont {Z.}~\bibnamefont {Wang}},\ }\href {\doibase 10.1088/1475-7516/2016/10/017} {\bibfield  {journal} {\bibinfo  {journal} {JCAP}\ }\textbf {\bibinfo {volume} {10}},\ \bibinfo {pages} {017} (\bibinfo {year} {2016})},\ \Eprint {http://arxiv.org/abs/1607.07872} {arXiv:1607.07872 [astro-ph.CO]} \BibitemShut {NoStop}%
\bibitem [{\citenamefont {Chen}\ \emph {et~al.}(2013)\citenamefont {Chen}, \citenamefont {Firouzjahi}, \citenamefont {Komatsu}, \citenamefont {Namjoo},\ and\ \citenamefont {Sasaki}}]{Chen:2013eea}%
  \BibitemOpen
  \bibfield  {author} {\bibinfo {author} {\bibfnamefont {X.}~\bibnamefont {Chen}}, \bibinfo {author} {\bibfnamefont {H.}~\bibnamefont {Firouzjahi}}, \bibinfo {author} {\bibfnamefont {E.}~\bibnamefont {Komatsu}}, \bibinfo {author} {\bibfnamefont {M.~H.}\ \bibnamefont {Namjoo}}, \ and\ \bibinfo {author} {\bibfnamefont {M.}~\bibnamefont {Sasaki}},\ }\href {\doibase 10.1088/1475-7516/2013/12/039} {\bibfield  {journal} {\bibinfo  {journal} {JCAP}\ }\textbf {\bibinfo {volume} {12}},\ \bibinfo {pages} {039} (\bibinfo {year} {2013})},\ \Eprint {http://arxiv.org/abs/1308.5341} {arXiv:1308.5341 [astro-ph.CO]} \BibitemShut {NoStop}%
\bibitem [{\citenamefont {Cai}\ \emph {et~al.}(2018)\citenamefont {Cai}, \citenamefont {Chen}, \citenamefont {Namjoo}, \citenamefont {Sasaki}, \citenamefont {Wang},\ and\ \citenamefont {Wang}}]{Cai:2018dkf}%
  \BibitemOpen
  \bibfield  {author} {\bibinfo {author} {\bibfnamefont {Y.-F.}\ \bibnamefont {Cai}}, \bibinfo {author} {\bibfnamefont {X.}~\bibnamefont {Chen}}, \bibinfo {author} {\bibfnamefont {M.~H.}\ \bibnamefont {Namjoo}}, \bibinfo {author} {\bibfnamefont {M.}~\bibnamefont {Sasaki}}, \bibinfo {author} {\bibfnamefont {D.-G.}\ \bibnamefont {Wang}}, \ and\ \bibinfo {author} {\bibfnamefont {Z.}~\bibnamefont {Wang}},\ }\href {\doibase 10.1088/1475-7516/2018/05/012} {\bibfield  {journal} {\bibinfo  {journal} {JCAP}\ }\textbf {\bibinfo {volume} {05}},\ \bibinfo {pages} {012} (\bibinfo {year} {2018})},\ \Eprint {http://arxiv.org/abs/1712.09998} {arXiv:1712.09998 [astro-ph.CO]} \BibitemShut {NoStop}%
\bibitem [{\citenamefont {Davies}\ \emph {et~al.}(2022)\citenamefont {Davies}, \citenamefont {Carrilho},\ and\ \citenamefont {Mulryne}}]{Davies:2021loj}%
  \BibitemOpen
  \bibfield  {author} {\bibinfo {author} {\bibfnamefont {M.~W.}\ \bibnamefont {Davies}}, \bibinfo {author} {\bibfnamefont {P.}~\bibnamefont {Carrilho}}, \ and\ \bibinfo {author} {\bibfnamefont {D.~J.}\ \bibnamefont {Mulryne}},\ }\href {\doibase 10.1088/1475-7516/2022/06/019} {\bibfield  {journal} {\bibinfo  {journal} {JCAP}\ }\textbf {\bibinfo {volume} {06}},\ \bibinfo {pages} {019} (\bibinfo {year} {2022})},\ \Eprint {http://arxiv.org/abs/2110.08189} {arXiv:2110.08189 [astro-ph.CO]} \BibitemShut {NoStop}%
\bibitem [{Note1()}]{Note1}%
  \BibitemOpen
  \bibinfo {note} {Strictly speaking, this interaction Hamiltonian is a function of redefined field $\protect \bm {\zeta }$, which was first introduced by Maldacena \cite {Maldacena:2002vr}. The relation between $\zeta $ and $\protect \bm {\zeta }$ is $\zeta = \protect \bm {\zeta }+ \protect \frac {1}{4}\eta \protect \bm {\zeta }^2 + \protect \frac {1}{H} \protect \dot {\protect \bm {\zeta }}\protect \bm {\zeta }+ \protect \dots $, where dots represent SR or a superhorizon suppressed term. It is shown that the field redefinition method is equivalent to considering boundary interaction \cite {Arroja:2011yj, Burrage:2011hd}. Moreover, quartic self-interaction \cite {Jarnhus:2007ia, Arroja:2008ga} with first-order perturbation theory is expected to generate independent contributions to the one-loop correction because it involves a higher-order SR parameter.}\BibitemShut {Stop}%
\bibitem [{\citenamefont {Kristiano}\ and\ \citenamefont {Yokoyama}()}]{Kristiano:2023scm}%
  \BibitemOpen
  \bibfield  {author} {\bibinfo {author} {\bibfnamefont {J.}~\bibnamefont {Kristiano}}\ and\ \bibinfo {author} {\bibfnamefont {J.}~\bibnamefont {Yokoyama}},\ }\href@noop {} {\ }\Eprint {http://arxiv.org/abs/2303.00341} {arXiv:2303.00341 [hep-th]} \BibitemShut {NoStop}%
\bibitem [{Note2()}]{Note2}%
  \BibitemOpen
  \bibinfo {note} {This can be derived by introducing physical wavenumber cutoff $\Lambda a(\tau )$ to the upper bound of the wavenumber integral in \protect \eqref {onel1} and \protect \eqref {onel2} \cite {Senatore:2009cf}. After performing a time integral \protect \eqref {tint}, the cutoff is evaluated at $\tau = \tau _e$, which leads to \protect \eqref {intlambda}}\BibitemShut {NoStop}%
\bibitem [{Note3()}]{Note3}%
  \BibitemOpen
  \bibinfo {note} {Dimensional regularization leads to a similar expression, except $\log \protect \tilde {\Lambda }$ is changed to $1/\delta $, where $\delta $ is a small correction to the spatial dimension. Another difference is that dimensional regularization cannot capture the quadratic divergence $\protect \tilde {\Lambda }^2$, as expected. Compared to one-loop computation for the case considered in \cite {Senatore:2009cf, delRio:2018vrj}, the time integral in our case is dominated by $\tau = \tau _e$, not the CMB horizon crossing time $\tau = -1/p$.}\BibitemShut {Stop}%
\bibitem [{\citenamefont {Armendariz-Picon}\ \emph {et~al.}(1999)\citenamefont {Armendariz-Picon}, \citenamefont {Damour},\ and\ \citenamefont {Mukhanov}}]{Armendariz-Picon:1999hyi}%
  \BibitemOpen
  \bibfield  {author} {\bibinfo {author} {\bibfnamefont {C.}~\bibnamefont {Armendariz-Picon}}, \bibinfo {author} {\bibfnamefont {T.}~\bibnamefont {Damour}}, \ and\ \bibinfo {author} {\bibfnamefont {V.~F.}\ \bibnamefont {Mukhanov}},\ }\href {\doibase 10.1016/S0370-2693(99)00603-6} {\bibfield  {journal} {\bibinfo  {journal} {Phys. Lett. B}\ }\textbf {\bibinfo {volume} {458}},\ \bibinfo {pages} {209} (\bibinfo {year} {1999})},\ \Eprint {http://arxiv.org/abs/hep-th/9904075} {arXiv:hep-th/9904075} \BibitemShut {NoStop}%
\bibitem [{\citenamefont {Kobayashi}\ \emph {et~al.}(2011)\citenamefont {Kobayashi}, \citenamefont {Yamaguchi},\ and\ \citenamefont {Yokoyama}}]{Kobayashi:2011nu}%
  \BibitemOpen
  \bibfield  {author} {\bibinfo {author} {\bibfnamefont {T.}~\bibnamefont {Kobayashi}}, \bibinfo {author} {\bibfnamefont {M.}~\bibnamefont {Yamaguchi}}, \ and\ \bibinfo {author} {\bibfnamefont {J.}~\bibnamefont {Yokoyama}},\ }\href {\doibase 10.1143/PTP.126.511} {\bibfield  {journal} {\bibinfo  {journal} {Prog. Theor. Phys.}\ }\textbf {\bibinfo {volume} {126}},\ \bibinfo {pages} {511} (\bibinfo {year} {2011})},\ \Eprint {http://arxiv.org/abs/1105.5723} {arXiv:1105.5723 [hep-th]} \BibitemShut {NoStop}%
\bibitem [{\citenamefont {Arroja}\ and\ \citenamefont {Tanaka}(2011)}]{Arroja:2011yj}%
  \BibitemOpen
  \bibfield  {author} {\bibinfo {author} {\bibfnamefont {F.}~\bibnamefont {Arroja}}\ and\ \bibinfo {author} {\bibfnamefont {T.}~\bibnamefont {Tanaka}},\ }\href {\doibase 10.1088/1475-7516/2011/05/005} {\bibfield  {journal} {\bibinfo  {journal} {JCAP}\ }\textbf {\bibinfo {volume} {05}},\ \bibinfo {pages} {005} (\bibinfo {year} {2011})},\ \Eprint {http://arxiv.org/abs/1103.1102} {arXiv:1103.1102 [astro-ph.CO]} \BibitemShut {NoStop}%
\bibitem [{\citenamefont {Burrage}\ \emph {et~al.}(2011)\citenamefont {Burrage}, \citenamefont {Ribeiro},\ and\ \citenamefont {Seery}}]{Burrage:2011hd}%
  \BibitemOpen
  \bibfield  {author} {\bibinfo {author} {\bibfnamefont {C.}~\bibnamefont {Burrage}}, \bibinfo {author} {\bibfnamefont {R.~H.}\ \bibnamefont {Ribeiro}}, \ and\ \bibinfo {author} {\bibfnamefont {D.}~\bibnamefont {Seery}},\ }\href {\doibase 10.1088/1475-7516/2011/07/032} {\bibfield  {journal} {\bibinfo  {journal} {JCAP}\ }\textbf {\bibinfo {volume} {07}},\ \bibinfo {pages} {032} (\bibinfo {year} {2011})},\ \Eprint {http://arxiv.org/abs/1103.4126} {arXiv:1103.4126 [astro-ph.CO]} \BibitemShut {NoStop}%
\bibitem [{\citenamefont {Jarnhus}\ and\ \citenamefont {Sloth}(2008)}]{Jarnhus:2007ia}%
  \BibitemOpen
  \bibfield  {author} {\bibinfo {author} {\bibfnamefont {P.~R.}\ \bibnamefont {Jarnhus}}\ and\ \bibinfo {author} {\bibfnamefont {M.~S.}\ \bibnamefont {Sloth}},\ }\href {\doibase 10.1088/1475-7516/2008/02/013} {\bibfield  {journal} {\bibinfo  {journal} {JCAP}\ }\textbf {\bibinfo {volume} {02}},\ \bibinfo {pages} {013} (\bibinfo {year} {2008})},\ \Eprint {http://arxiv.org/abs/0709.2708} {arXiv:0709.2708 [hep-th]} \BibitemShut {NoStop}%
\bibitem [{\citenamefont {Arroja}\ and\ \citenamefont {Koyama}(2008)}]{Arroja:2008ga}%
  \BibitemOpen
  \bibfield  {author} {\bibinfo {author} {\bibfnamefont {F.}~\bibnamefont {Arroja}}\ and\ \bibinfo {author} {\bibfnamefont {K.}~\bibnamefont {Koyama}},\ }\href {\doibase 10.1103/PhysRevD.77.083517} {\bibfield  {journal} {\bibinfo  {journal} {Phys. Rev. D}\ }\textbf {\bibinfo {volume} {77}},\ \bibinfo {pages} {083517} (\bibinfo {year} {2008})},\ \Eprint {http://arxiv.org/abs/0802.1167} {arXiv:0802.1167 [hep-th]} \BibitemShut {NoStop}%
\end{thebibliography}%

\end{document}